\documentstyle[12pt,epsfig,axodraw]{article}
\begin{document}
\pagestyle{plain} \setcounter{page}{1} \baselineskip=0.3in
\begin{titlepage}
\vspace{.5cm}

\begin{center}
{\Large Supersymmetric Electroweak Corrections to Heavier Top
Squark Decay into  Lighter Top Squark and  Neutral Higgs Boson}

\vspace{.2in} Qiang Li, Li Gang Jin  and  Chong Sheng Li \\
\vspace{.2in} Department of Physics, Peking University, Beijing
100871, People's Republic of China \\
\end{center}
\vspace{.4in}
\begin{footnotesize}
\begin{center}
\begin{minipage}{5in}
\baselineskip=0.25in
\begin{center} ABSTRACT \end{center}
We calculate the Yukawa corrections of order ${\cal
O}(\alpha_{ew}m_{t(b)}^2/m_W^2)$, ${\cal
O}(\alpha_{ew}m_{t(b)}^3/m_W^3)$ and ${\cal
O}(\alpha_{ew}m_{t(b)}^4/m_W^4)$ to the widths of the decays
$\tilde t_2\rightarrow \tilde t_1$+$(h^0,H^0,A^0)$ in the Minimal
Supersymmetric Standard Model,   and perform a detailed numerical
analysis. We also compare the results with the ones presented in
an earlier literature, where  the ${\cal O}(\alpha_{s})$ SUSY-QCD
corrections to the same three decay processes have been
calculated.
 Our numerical results show that for the
decays $\tilde t_2\rightarrow \tilde t_1$+$h^{0}$ , $\tilde
t_2\rightarrow \tilde t_1$+$H^{0}$, the Yukawa corrections are
significant in most of the parameter range, which can reach a few
ten percent, and for the decay $\tilde t_2\rightarrow\tilde
t_1$+$A^{0}$, the Yukawa corrections are relatively smaller, which
are only a few percent. The numerical calculations also show that
using the running quark masses and the running trilinear coupling
$A_t$ , which include the QCD, SUSY-QCD, SUSY-Electroweak effects
and resume all high order ($\tan\beta$)-enhanced effects, can
vastly improve the convergence of the perturbation expansion. We
also discuss the effects of the running of the higgsino mass
parameter $\mu$ on the corrections, and find that they are
significant, too, especially for large $\tan\beta$.
\end{minipage}\end{center}
\end{footnotesize}
\vfill

PACS number(s): 14.80.Cp; 14.80.Ly; 12.38.Bx
\end{titlepage}
\eject \baselineskip=0.3in
\section{Introduction}
Incorporation of supersymmetry is one of the most attractive and
promising possibilities for new physics beyond the Standard Model
(SM)\cite{MSSM,Haber}, and the Minimal Supersymmetric Standard
Model (MSSM) is a popular candidate for new physics in this way.
In the MSSM there are many new particles. For example, every quark
has two spin zero partners called squarks $\tilde {q}_L$ and
$\tilde {q}_R$,  one for each chirality eigenstate, which mix to
form the mass eigenstates $\tilde {q}_1$ and $\tilde {q}_2$. For
the third generation quarks, due to  large Yukawa couplings, there
may be large mass differences between the lighter mass eigenstate
and the heavier one, which implies in general a very complex decay
pattern of the heavier state.

As we know, the next generation of colliders, such as the Large
Hadron Collider (LHC), the upgraded Tevatron, $e^+e^-$ linear
colliders, and $\mu^+\mu^-$ collider will push the discovery reach
for supersymmetric (SUSY) paticles with masses up to 2.5
TeV\cite{LHC,LC} and allow for precise measurement of the MSSM
parameters. Thus a more accurate calculations of the deacy
mechanisms beyond the tree level are necessary. The dominate decay
modes of the heavier squarks are shown as below:
$$\tilde {t}_i\rightarrow t\tilde {x}_k^0,b\tilde {x}_k^+;
 \tilde {b}_i\rightarrow b\tilde {x}_k^0,t\tilde {x}_k^+; \ \
 \tilde {t}_i\rightarrow t\tilde {g};
 \tilde {b}_i\rightarrow b\tilde {g} ; $$
$$\tilde {t}_2\rightarrow \tilde {t}_1 Z^0;
 \tilde {t}_i\rightarrow \tilde {b}_j W^+; \ \
 \tilde {b}_2\rightarrow \tilde {b}_1 Z^0;
 \tilde {b}_i\rightarrow \tilde {t}_j W^-;$$
$$ \tilde {b}_i\rightarrow \tilde {t}_j H^-;
 \tilde {t}_i\rightarrow \tilde {b}_j H^+;\  \ \
\tilde {t}_2\rightarrow \tilde {t}_1 (h^0,H^0,A^0).$$ All these
squark decays have been extensively discussed at the
tree-level\cite{treelevel,Bartl,Hidaka}. Up to now, one-loop QCD
and supersymmetric (SUSY) QCD corrections to above decay channels
have been calculated too \cite{Bartl,QCDcorrection,QCDC}, while
the Yukawa corrections and the full electroweak one-loop
radiative corrections to the squark decays into quarks plus
charginos/neutralinos were given in Ref.\cite{Yukawacorrection}
and Ref.\cite{full}, respectively. Also the Yukawa corrections to
the squark decays into quarks plus gluinos were given in
Refs.\cite{Ma,Yang}, and the Yukawa corrections to the heavier
squark decays into  lighter squarks plus vetor bosons were given
in Ref.\cite{vetor}. Recently,  the Yukawa corrections to the
bottom squark decays into lighter top squarks plus  charged Higgs
bosons has been presented in Ref.\cite{LiGang}. So only the
electroweak radiative corrections to the heavier top squark decays
into lighter top squarks plus neutral Higgs bosons have not been
calculated yet, including the Yukawa corrections to these
processes.

In this paper, we present the calculations of the Yukawa
corrections of order ${\cal O}(\alpha_{ew}m_{t(b)}^2/m_W^2)$,
${\cal O}(\alpha_{ew}m_{t(b)}^3/m_W^3)$, and ${\cal
O}(\alpha_{ew}m_{t(b)}^4/m_W^4)$ to the widths of the heavier top
squark decays into  lighter top squarks plus  neutral Higgs
 bosons, i.e.the decays $\tilde {t}_2 \rightarrow \tilde {t}_1 +
(h^0,H^0,A^0)$. These corrections are mainly induced by the Yukawa
couplings from Higgs-quark-quark couplings, Higgs-squark-squark
couplings, Higgs-Higgs-squark-squark couplings,
chargino(neutralino)-quark-squark couplings, and
squark-squark-squark-squark couplings. As shown in Ref.\cite{RAt},
the Higgs-Squark-Squark couplings receive large radiative
corrections, which can make the perturbation calculation of the
relevant Squark or Higgs boson decay widths quite unreliable in
some region of the parameter space.
 When the correction term is negative,
the corrected width can even become negative, which clearly makes
no sense.  In order to solve this problem, we use the running
quark masses and the running trilinear coupling $A_t$ \cite{RAt},
and vastly improve the convergence of the perturbation expansion.
We also discuss the effects of the running of the higgsino mass
parameter $\mu$ on the corrections, and find that they are
significant, too, especially for large $\tan\beta$.
\section{Notation and tree-level result}

In order to make this paper self-contained, we first summarize our
notation and present the relevant interaction Lagrangians of the
MSSM and the tree-level decay rates for $\tilde {t}_2 \rightarrow
\tilde {t}_1+ (h^{0},H^{0},A^{0})$.

The current eigenstates $\tilde{q}_L$ and $\tilde{q}_R$ are
related to the mass eigenstates $\tilde{q}_1$ and $\tilde{q}_2$ by
\begin{equation}
\left(\begin{array}{c} \tilde{q}_1 \\ \tilde{q}_2 \end{array}
\right)= R^{\tilde{q}}\left(\begin{array}{c} \tilde{q}_L \\
\tilde{q}_R \end{array} \right), \ \ \ \ \
R^{\tilde{q}}=\left(\begin{array}{cc} \cos\theta_{\tilde{q}} &
\sin\theta_{\tilde{q}} \\ -\sin\theta_{\tilde{q}} &
\cos\theta_{\tilde{q}}
\end{array} \right)
\end{equation}
with $0 \leq \theta_{\tilde{q}} < \pi$ by convention.
Correspondingly, the mass eigenvalues $m_{\tilde{q}_1}$ and
$m_{\tilde{q}_2}$ (with $m_{\tilde{q}_1}\leq m_{\tilde{q}_2}$) are
given by
\begin{eqnarray}\label{Mq2}
\left(\begin{array}{cc} m_{\tilde{q}_1}^2 & 0 \\ 0 &
m_{\tilde{q}_2}^2 \end{array} \right)=R^{\tilde{q}}
M_{\tilde{q}}^2 (R^{\tilde{q}})^\dag, \ \ \ \ \
M_{\tilde{q}}^2=\left(\begin{array}{cc} m_{\tilde{q}_L}^2 & a_qm_q
\\ a_qm_q & m_{\tilde{q}_R}^2 \end{array} \right)
\end{eqnarray}
with
\begin{eqnarray}
m^2_{\tilde{q}_L} &=& M^2_{\tilde{Q}} +m_q^2
+m_Z^2\cos2\beta(I_{3L}^q -e_q\sin^2\theta_W), \\
m^2_{\tilde{q}_R} &=& M^2_{\{\tilde{U},\tilde{D}\}} +m_q^2
+m_Z^2\cos2\beta e_q\sin^2\theta_W, \\
a_q &=& A_q -\mu\{\cot\beta, \tan\beta\}
\end{eqnarray}
for \{up, down\} type squarks. Here $M_{\tilde{q}}^2$ is the squark
mass matrix. $M_{\tilde{Q},\tilde{U},\tilde{D}}$ and $A_{t,b}$ are
soft SUSY-breaking parameters and $\mu$ is the higgsino
mass parameter . $I_{3L}^q$ and $e_q$ are
the third component of the weak isospin and the electric charge of
the quark $q$, respectively.

Defining $H_k=(h^0, H^0, A^0, G^0, H^{\pm}, G^{\pm})$ ($k$=1,...,6),
one can write the relevant lagrangian density in the
($\tilde{q}_1,\tilde{q}_2$) basis as following form ($i,j$=1,2;
$\alpha$ and $\beta$ are flavor indices):
\begin{eqnarray}\label{lagrangian}
&& {\cal L}_{\rm relevant}=H_k\bar{q}^{\beta}(a_k^{\alpha}P_L
+b_k^{\alpha}P_R)q^{\alpha}
+(G^{\tilde{\alpha}}_k)_{ij}H_k\tilde{q}_j^{\beta\ast}\tilde{q}^\alpha_i
+g\bar{q}(a_{ik}^{\tilde{q}}P_R
+b_{ik}^{\tilde{q}})\tilde{\chi}^0_k\tilde{q}_i \nonumber \\
&& \hspace{1.5cm} +g\bar{t}(l_{ik}^{\tilde{b}}P_R
+k_{ik}^{\tilde{b}}P_L)\tilde{\chi}^+_k\tilde{b}_i
+g\bar{b}(l_{ik}^{\tilde{t}}P_R
+k_{ik}^{\tilde{t}}P_L)\tilde{\chi}_k^{+c}\tilde{t}_i \nonumber \\
&& \hspace{1.5cm} +(G^{\tilde{\alpha}}_{lk})_{ij}H_lH_k
\tilde{q}_j^{\beta\ast}\tilde{q}^\alpha_i +h.c.,
\end{eqnarray}
with\begin{eqnarray} (G^{\tilde{\alpha}}_{lk})_{ij}
=[R^{\tilde{\alpha}} \hat{G}_{lk}^{\tilde{\alpha}}
(R^{\tilde{\beta}})^T]_{ij} \hspace{0.3cm} (l,k=1,...,6)
\end{eqnarray}\begin{eqnarray}
(G^{\tilde{\alpha}}_{k})_{ij} =[R^{\tilde{\alpha}}
\hat{G}_{k}^{\tilde{\alpha}} (R^{\tilde{\beta}})^T]_{ij}\hspace{0.3cm} (k=1,...,6)
\end{eqnarray}

\noindent where $\hat{G}^{\tilde{\alpha}}_{k} $ and $
\hat{G}^{\tilde{\alpha}}_{lk}$ are the couplings in the
($\tilde{q}_L,\tilde{q}_R$) basis ,  and their explicit forms are
shown in Appendix A. The notations $a_k^\alpha$, $b_k^\alpha$
(k=1,...,6), and $a_{ik}^{\tilde{q}}$, $b_{ik}^{\tilde{q}}$
(k=1,...,4), and $l_{ik}^{\tilde{q}}$, $k_{ik}^{\tilde{q}}$
(k=1,2) used in Eq.(\ref{lagrangian}) are defined also in Appendix
A.

The tree-level amplitudes of the three decay processes, as shown in
Fig.1(a), are given by

\begin{eqnarray}
M^{(0)}_1=i[R^{\tilde{t}} \left(\begin{array}{cc}
-\sqrt2m_{t}h_{t}\cos\alpha+\frac{gm_{z}\sin(\alpha+\beta)}{c_W}C_{tL}
& \frac{-h_t}{\sqrt2}(A_t\cos\alpha+\mu\sin\alpha) \\
\frac{-h_t}{\sqrt2}(A_t\cos\alpha+\mu\sin\alpha)&
-\sqrt2m_{t}h_{t}\cos\alpha+\frac{gm_{z}\sin(\alpha+\beta)}{c_W}C_{tR}
\end{array} \right)
 (R^{\tilde{t}})^T]_{21}
\end{eqnarray}

\noindent for  $\tilde{t}_2 \rightarrow \tilde{t}_1 h^{0}$,

\begin{eqnarray}
M^{(0)}_2=i[R^{\tilde{t}} \left(\begin{array}{cc}
-\sqrt2m_{t}h_{t}\sin\alpha-\frac{gm_{z}\cos(\alpha+\beta)}{c_W}C_{tL}
& \frac{-h_t}{\sqrt2}(A_t\sin\alpha-\mu\cos\alpha) \\
\frac{-h_t}{\sqrt2}(A_t\sin\alpha-\mu\cos\alpha)&
-\sqrt2m_{t}h_{t}\sin\alpha-\frac{gm_{z}\cos(\alpha+\beta)}{c_W}C_{tR}
\end{array} \right)
 (R^{\tilde{t}})^T]_{21}
\end{eqnarray}

\noindent for  $\tilde{t}_2 \rightarrow \tilde{t}_1 H^{0}$,  and

\begin{eqnarray}
\noindent M^{(0)}_3=\frac{gm_{t}}{2m_{W}}[R^{\tilde{t}}
\left(\begin{array}{cc}
0& A_t\cot\beta+\mu \\
-A_t\cot\beta-\mu& 0
\end{array} \right)
 (R^{\tilde{t}})^T]_{21}
\end{eqnarray}
for  $\tilde{t}_2 \rightarrow \tilde{t}_1 A^{0}$.
 Here
$h_{t}=\frac{gm_{t}}{\sqrt2m_{W}\sin\beta}$,
$h_{b}=\frac{gm_{b}}{\sqrt2m_{W}\cos\beta}$,
$C_{tL}=I_{3L}^{t}-e_{t}s^2_W$, $C_{tR}=e_{t}s^2_w$,
$s_W=\sin\theta_w$, and $c_w=\cos\theta_w$.
 $I_{3L}^{t}=\frac{1}{2}$  and  $e_{t}=\frac{2}{3}$   for the top
squark, $I_{3L}^{b}=-\frac{1}{2}$  and  $ e_{b}=-\frac{1}{3}$  for
the bottom squark. The tree-level decay width is thus given by
\begin{eqnarray}
\Gamma^{(0)}_s =\frac{|M^{(0)}_s|^2\lambda^{1/2}
(m_{\tilde{t}_2}^2, m_{\tilde{t}_1}^2, m_{H^0_{s}}^2)}{16\pi
m_{\tilde{t}_2}^3},
\end{eqnarray}
where $\lambda(x,y,z)=(x-y-z)^2-4yz$ and s=(1,2,3) corresponds to
the decay into $(h^{0},H^{0},A^{0})$, respectively.
\section{Yukawa corrections}

The Feynman diagrams contributing to the Yukawa corrections to
$\tilde{t}_2 \rightarrow \tilde{t}_1 H^0_i$ are shown in
Figs.1(b)--(f) and Fig.2. We carried out the calculation in the
t'Hooft-Feynman gauge and used the dimensional reduction, which
preserves supersymmetry, for regularization of the ultraviolet
divergences in the virtual loop corrections using the
on-mass-shell renormalization scheme\cite{on-mass}, in which the
fine-structure constant $\alpha_{ew}$ and physical masses are
chosen to be the renormalized parameters, and finite parts of the
counterterms are fixed by the renormalization conditions. The
coupling constant $g$ is related to the input parameters $e$,
$m_W$ and $m_Z$ via $g^2=e^2/s_w^2$ and $s_w^2=1-m_W^2/m_Z^2$. As
for the renormalization of the parameters in the Higgs sector and
the squark sector, it will be described in detail below.

The relevant renormalization constants are defined as
\begin{eqnarray}
&& m_{W0}^2=m_W^2 +\delta m_W^2, \hspace{0.5cm} m_{Z0}^2=m_Z^2
+\delta m_Z^2, \nonumber \\ && m_{q0}=m_q +\delta m_q,
\hspace{1.0cm} m_{\tilde{q}_i0}^2 =m_{\tilde{q}_i}^2 +\delta
m_{\tilde{q}_i}^2, \nonumber \\ && A_{q0}=A_q +\delta A_q,
\hspace{1.2cm} \mu_0 =\mu +\delta\mu, \nonumber \\ &&
\theta_{\tilde{q}0}=\theta_{\tilde{q}} +\delta\theta_{\tilde{q}},
\hspace{1.5cm} \tan\beta_0=(1 +\delta Z_\beta) \tan\beta,
\nonumber \\ && \sin\alpha_0=(1 +\delta Z_\alpha) \sin\alpha,
\nonumber \\&& \tilde{q}_{i0}=(1 +\delta Z^{\tilde{q}}_i)^{1/2}
+\delta Z^{\tilde{q}}_{ij}\tilde{q}_j, \nonumber \\ && H^0_0=(1
+\delta Z_{H^0})^{1/2}H^0 +\delta Z_{H^0h^0}h^0,  \nonumber
\\ && h^0_0=(1
+\delta Z_{h^0})^{1/2}h^0 +\delta Z_{h^0H^0}H^0,  \nonumber\\ &&
G_0^-=(1 +\delta Z_{G^-})^{1/2}G^- +\delta Z_{GH}H^-,\nonumber\\ &&
A_0^0=(1 +\delta Z_{A^0})^{1/2}A^0
\end{eqnarray}
with $q=t,b$. Here we introduce the mixing of $H^-$ and
$G^-$\cite{GH}.

Taking into account the Yukawa corrections, the renormalized
amplitude for $\tilde{t}_2 \rightarrow \tilde{t}_1 H_s^0$ is given
by
\begin{equation}
M^{ren}_s=M^{(0)}_s +\delta M^{(v)}_s +\delta M^{(c)}_s,
\end{equation}
where $\delta M^{(v)}_s$ and $\delta M^{(c)}_s$ are the vertex
corrections and  the counterterms, respectively.

The calculations of the vertex corrections from Fig.1(b)-1(f)
result in
\begin{eqnarray}
&&\delta M^{(v)}_{s=1,2,3}=
  \frac{i}{16\pi^2}\sum_{k=1}^6\sum_j(G_{sk}^{\tilde{t}})_{2j}
  (G_{k}^{\tilde{q}})_{j1}B_0(m^2_{\tilde{t}_1},m_{H_k^0},m_{\tilde{q}_j})
  \nonumber \\
  && \hspace{1.6cm}
 -\frac{i}{16\pi^2}\sum_{k=1}^6\sum_{ij}(G_{s}^{\tilde{t}})_{ij}
  (G_{k}^{\tilde{t}})_{2i}(G_{k}^{\tilde{q}})_{j1}
  C_0(p_{\tilde{t}_1},p_{H_s^0},m_{H_k^0},m_{\tilde{q}_j},m_{\tilde{q}_i})
  \nonumber \\
  && \hspace{1.6cm}
  +\frac{i}{16\pi^2}\sum_{k=1}^6\sum_j(G_{sk}^{\tilde{t}})_{j1}
  (G_{k}^{\tilde{q}})_{2j}B_0(m^2_{\tilde{t}_2},
  m_{H_k^0},m_{\tilde{q}_j})
  \nonumber \\
   && \hspace{1.6cm}
  -\frac{i}{16\pi^2}\sum_{ij}\sin\theta_{\tilde{t}}\cos\theta_{\tilde{t}}
  (h_t^2R^{\tilde b}_{i1}
  R^{\tilde{b}}_{j1}-h_b^2R^{\tilde{b}}_{i2}R^{\tilde{b}}_{j2})
(G_{s}^{\tilde{b}})_{ij}B_0(m^2_{H_s^0},m_{\tilde{b}_j},m_{\tilde{b}_i})
   \nonumber \\
   && \hspace{1.6cm}
-\frac{ih_t^2}{16\pi^2}\{\hspace{0.15cm}
[3(\sin^4\theta_{\tilde{t}}+\cos^4\theta_{\tilde{t}})
-2\sin^2\theta_{\tilde{t}}\cos^2\theta_{\tilde{t}}]
(G_{s}^{\tilde{t}})_{21}B_0(m^2_{H_s^0},m_{\tilde{t}_1},m_{\tilde{t}_2})
  \nonumber \\
   && \hspace{2.5cm}
  -8\sin^2\theta_{\tilde{t}}\cos^2\theta_{\tilde{t}}
(G_{s}^{\tilde{t}})_{12}B_0(m^2_{H_s^0},m_{\tilde{t}_2},m_{\tilde{t}_1})
 \nonumber \\
   && \hspace{2.5cm}
  +4\sin\theta_{\tilde{t}}\cos\theta_{\tilde{t}}\cos2\theta_{\tilde{t}}
(G_{s}^{\tilde{t}})_{11}B_0(m^2_{H_s^0},m_{\tilde{t}_1},m_{\tilde{t}_1})
\nonumber \\
&& \hspace{2.5cm}
-4\sin\theta_{\tilde{t}}\cos\theta_{\tilde{t}}\cos2\theta_{\tilde{t}}
(G_{s}^{\tilde{t}})_{22}
B_0(m^2_{H_s^0},m_{\tilde{t}_2},m_{\tilde{t}_2})
 \hspace{0.15cm}\}
   \nonumber \\
   && \hspace{1.6cm}
   +F_{\chi s}.
\end{eqnarray}
where $F_{\chi s}$ is the remains, which are given by
\begin{eqnarray}
&&\hspace{0.8cm} F_{\chi s}= -\frac{ig^2
h_t\cos(\alpha-(s-1)\frac{\pi}{2})}{16\sqrt{2}\pi^2}\sum_i\{\hspace{0.3cm}
[2(a_{2i}^{\tilde t}{a_{1i}^{\tilde t}}^*+b_{2i}^{\tilde
t}{b_{1i}^{\tilde t}}^*)
m_t][(2m_{\tilde{t}_1}^2+p_{\tilde{t}_1}p_{H_s^0})C_{11}
\nonumber \\
&& \hspace{1.2cm}
+(2p_{\tilde{t}_1}p_{H_s^0}+m_{H_s^0}^2)C_{12}
+2m_{\tilde{t}_1}^2C_{21}+2m_{H_s^0}^2C_{22}+4p_{\tilde{t}_1}p_{H_s^0}C_{23}+8C_{24}]
\nonumber \\
&& \hspace{1.2cm}
+[2(a_{2i}^{\tilde t}{b_{1i}^{\tilde t}}^*+b_{2i}^{\tilde t}{a_{1i}^{\tilde t}}^*)
m_{{\tilde\chi_{i}^0}}][(2m_{\tilde{t}_1}^2+p_{\tilde{t}_1}p_{H_s^0})C_{11}
+(2p_{\tilde{t}_1}p_{H_s^0}+m_{H_s^0}^2)C_{12}
\nonumber \\
   && \hspace{1.2cm}
+(m_{\tilde{t}_1}^2C_{21}+m_{H_s^0}^2C_{22}+2p_{\tilde{t}_1}p_{H_s^0}C_{23}
+4C_{24}+(m_{\tilde{t}_1}^2+p_{\tilde{t}_1}p_{H_s^0})C_0+m_t^2C_0]
\nonumber \\
  && \hspace{1.2cm}
   \}(p_{\tilde{t}_1},p_{H_s^0},m_{{\tilde\chi_{i}^0}},m_t,m_t)
   \nonumber \\
   && \hspace{1.2cm}
+\frac{ig^2 h_b\sin(\alpha-(s-1)\frac{\pi}{2})}{16\sqrt{2}\pi^2}\sum_i\{\hspace{0.3cm}
[2(l_{2i}^{\tilde t}{l_{1i}^{\tilde t}}^*+k_{2i}^{\tilde t}{k_{1i}^{\tilde t}}^*)
m_b][(2m_{\tilde{t}_1}^2+p_{\tilde{t}_1}p_{H_s^0})C_{11}
\nonumber \\
&& \hspace{1.2cm}
+(2p_{\tilde{t}_1}p_{H_s^0}+m_{H_s^0}^2)C_{12}
+2m_{\tilde{t}_1}^2C_{21}+2m_{H_s^0}^2C_{22}+4p_{\tilde{t}_1}p_{H_s^0}C_{23}+8C_{24}]
\nonumber \\
   && \hspace{1.2cm}
+[2(l_{2i}^{\tilde t}{k_{1i}^{\tilde t}}^*+k_{2i}^{\tilde t}{l_{1i}^{\tilde t}}^*)
m_{{\tilde\chi_{i}}^-}][(2m_{\tilde{t}_1}^2+p_{\tilde{t}_1}p_{H_s^0})C_{11}
+(2p_{\tilde{t}_1}p_{H_s^0}+m_{H_s^0}^2)C_{12}
\nonumber \\
   && \hspace{1.2cm}
+(m_{\tilde{t}_1}^2C_{21}+m_{H_s^0}^2C_{22}+2p_{\tilde{t}_1}p_{H_s^0}C_{23}
+4C_{24}+(m_{\tilde{t}_1}^2+p_{\tilde{t}_1}p_{H_s^0})C_0+m_b^2C_0]
\nonumber \\
  && \hspace{1.2cm}
   \}(p_{\tilde{t}_1},p_{H_s^0},m_{{\tilde\chi_{i}}^-},m_b,m_b)
 \end{eqnarray}
for s=(1,2), and
\\
\begin{eqnarray}
&&F_{\chi 3}= -\frac{g^3 m_t\cot\beta}{32\pi^2
m_W}\sum_i\{\hspace{0.3cm} [2(a_{2i}^{\tilde t}{a_{1i}^{\tilde
t}}^*-b_{2i}^{\tilde t}{b_{1i}^{\tilde t}}^*)
m_t](m_{A^0}^2C_{12}+p_{\tilde{t}_1}p_{A^0}C_{11})
\nonumber \\
&& \hspace{1.2cm}
+[2(a_{2i}^{\tilde t}{b_{1i}^{\tilde t}}^*-b_{2i}^{\tilde t}{a_{1i}^{\tilde t}}^*)
m_{{\tilde\chi_{i}^0}}][(2m_{\tilde{t}_1}^2+p_{\tilde{t}_1}p_{A^0})C_{11}
+(2p_{\tilde{t}_1}p_{A^0}+m_{A^0}^2)C_{12}
\nonumber \\
   && \hspace{1.2cm}
+m_{\tilde{t}_1}^2C_{21}+m_{A^0}^2C_{22}+2p_{\tilde{t}_1}p_{A^0}C_{23}
+4C_{24}+(p_{\tilde{t}_1}^2+p_{\tilde{t}_1}p_{A^0})C_0-m_t^2C_0]
\nonumber \\
  && \hspace{1.2cm}
   \}(p_{\tilde{t}_1},p_{A^0},m_{{\tilde\chi_{i}^0}},m_t,m_t)
   \nonumber \\
   && \hspace{1.2cm}
-\frac{ig^3 m_b\tan\beta}{32\pi^2 m_W}\sum_i\{\hspace{0.3cm}
[2(l_{2i}^{\tilde t}{l_{1i}^{\tilde t}}^*-k_{2i}^{\tilde
t}{k_{1i}^{\tilde t}}^*)
m_b](m_{A^0}^2C_{12}+p_{\tilde{t}_1}p_{A^0}C_{11})
\nonumber \\
   && \hspace{1.2cm}
+[2(l_{2i}^{\tilde t}{k_{1i}^{\tilde t}}^*-k_{2i}^{\tilde t}{l_{1i}^{\tilde t}}^*)
m_{{\tilde\chi_{i}}^-}][(2m_{\tilde{t}_1}^2+p_{\tilde{t}_1}p_{A^0})C_{11}
+(2p_{\tilde{t}_1}p_{A^0}+m_{A^0}^2)C_{12}
\nonumber \\
   && \hspace{1.2cm}
+m_{\tilde{t}_1}^2C_{21}+m_{A^0}^2C_{22}+2p_{\tilde{t}_1}p_{A^0}C_{23}+4C_{24}
+(p_{\tilde{t}_1}^2+p_{\tilde{t}_1}p_{A^0})C_0-m_b^2C_0]
\nonumber \\
  && \hspace{1.2cm}
   \}(p_{\tilde{t}_1},p_{A^0},m_{{\tilde\chi_{i}}^-},m_b,m_b)
 \end{eqnarray}
for s=3.
In above expressions $B_0$ and $C_{i(j)}$ are two- and three-point Feynman
integrals\cite{ABCD}, respectively.
For $q=t$, we have $k=1...4$.
For $q=b$, we have $k=5,6$.

The counterterms can be expressed as
\begin{eqnarray}
&& \delta M^{(c)}_1=i(G_1^{\tilde t})_{21}[ \frac{1}{2}(\delta Z_1+\delta Z_2+
\delta Z_{h^0})-2\tan2\theta_{\tilde t}\delta \theta_{\tilde t}+\frac{\delta g}{g}
-\frac{\delta m_W^2}{2m_W^2}-\cos^2\beta\delta \beta]\nonumber \\
  && \hspace{1.5cm}
-i\frac{gm_tA_t\cos\alpha}{2m_W\sin\beta} \cos2\theta_{\tilde t}
[\frac{A_t\delta m_t+m_t\delta A_t}{m_tA_t}+\tan^2\alpha\delta
Z_{\alpha}]
\nonumber \\
 && \hspace{1.5cm}
-i\frac{gm_t\mu\sin\alpha}{2m_W\sin\beta} \cos2\theta_{\tilde t}
[\frac{\delta\mu}{\mu}+\frac{\delta m_t}{m_t}-\delta Z_{\alpha}]
\nonumber \\
 && \hspace{1.5cm}
+i(G_1^{\tilde t})_{11}\delta Z_{12}+i(G_1^{\tilde t})_{22}\delta
Z_{21} +i(G_2^{\tilde t})_{21}\delta Z_{H^0h^0} ,
\\
&& \delta M^{(c)}_2=i(G_2^{\tilde t})_{21}[ \frac{1}{2}(\delta Z_1+\delta Z_2+
\delta Z_{H^0})-2\tan2\theta_{\tilde t}\delta \theta_{\tilde t}+\frac{\delta g}{g}
-\frac{\delta m_W^2}{2m_W^2}-\cos^2\beta\delta \beta]\nonumber \\
  && \hspace{1.5cm}
-i\frac{gm_tA_t\sin\alpha}{2m_W\sin\beta} \cos2\theta_{\tilde t}
[\frac{A_t\delta m_t+m_t\delta A_t}{m_tA_t}+\delta Z_{\alpha}]
\nonumber \\
 && \hspace{1.5cm}
+i\frac{g\mu m_t\cos\alpha}{2m_W\sin\beta} \cos2\theta_{\tilde t}
[\frac{\delta\mu}{\mu}+\frac{\delta m_t}{m_t}-\tan^2\alpha\delta
Z_{\alpha}]
\nonumber \\
 && \hspace{1.5cm}
+i(G_2^{\tilde t})_{11}\delta Z_{12}+i(G_2^{\tilde t})_{22}\delta
Z_{21} +i(G_1^{\tilde t})_{21}\delta Z_{h^0H^0} ,
\\
&& \delta M^{(c)}_3=i(G_3^{\tilde t})_{21}[ \frac{1}{2}(\delta Z_1+\delta Z_2+
\delta Z_{A^0})+\frac{\delta g}{g}
-\frac{\delta m_W^2}{2m_W^2}]\nonumber \\
  && \hspace{1.5cm}
-\frac{gm_tA_t\cos\alpha}{2m_W} \cot\beta [\frac{\delta(
m_tA_t)}{m_tA_t}-\delta Z_\beta] -\frac{g\mu m_t}{2m_W}
[\frac{\delta\mu}{\mu}+\frac{\delta m_t}{m_t}]  .
\end{eqnarray}
\noindent Here we consider only the counterterms from the Yukawa
couplings, and the explicit expressions of some renormalization
constants calculated from the self-energy diagrams in Fig.2 are
given in Appendix B. Other renormalization constants are fixed as
follows.

For $\delta Z_{GH}$, using the approach discussed in the two-Higgs
doublet model (2HDM) in \cite{GH}, we derived below its expression
in the MSSM, where the version of the Higgs potential is different
from one of Ref.\cite{GH}. First, the one-loop renormalized
two-point function is given by
\begin{eqnarray}
i\Gamma_{GH}(p^2)=i(p^2 -m_{H^-}^2)\delta Z_{HG} +ip^2\delta
Z_{GH} -iT_{GH} +i\Sigma_{GH}(p^2),
\end{eqnarray}
where $T_{GH}$ is the tadpole function, which is given by
\begin{equation}
T_{GH}=\frac{g}{2m_W}[T_{H_2}\sin(\alpha -\beta)
+T_{H_1}\cos(\alpha -\beta)].
\end{equation}
Next, from the on-shell renormalization condition, we obtained
\begin{equation}
\delta Z_{GH}=\frac{1}{m_{H^-}^2}[T_{GH} -\Sigma_{GH}(m_{H^-}^2)].
\end{equation}
The explicit expressions of $\Sigma_{GH}$ and the tadpole
counterterms $T_{H_k}$ $(k=1,2)$ are given in Appendix B.

For the renormalization of the parameter $\beta$, following the
analysis of Ref.\cite{Mendez}, we fixed the renormalization
constant by the requirement that the on-mass-shell $H^+ \bar{\tau}
\nu_\tau$ coupling remain the same form as in Eq.(3) of
Ref.\cite{Mendez} to all orders of perturbation theory. However,
with introducing the mixing of $H^-$ and $G^-$ instead of $H^-$
and $W^-$, the expression of $\delta Z_\beta$ is then changed to
\begin{eqnarray}
&& \delta Z_\beta=\frac{1}{2}\frac{\delta
  m_W^2}{m_W^2} -\frac{1}{2}\frac{\delta m_Z^2}{m_Z^2}
  +\frac{1}{2}\frac{\delta m_Z^2 -\delta m_W^2}{m_Z^2 -m_W^2}
  -\frac{1}{2}\delta Z_{H^+} +\cot\beta\delta Z_{GH}.
\end{eqnarray}

For the counterterm of squark mixing angle $\theta_{\tilde{q}}$,
using the same renormalized scheme as Ref.\cite{Yukawacorrection},
one has
\begin{equation}
\delta\theta_{\tilde{q}} =\frac{{\rm
Re}[\Sigma_{12}^{\tilde{q}}(m_{\tilde{q}_1}^2)
+\Sigma_{12}^{\tilde{q}}(m_{\tilde{q}_2}^2)]}{2(m_{\tilde{q}_1}^2
-m_{\tilde{q}_2}^2)},
\end{equation}
 where the explicit expressions of the $\Sigma_{ij}$ functions  arising
 from the self-energy diagrams due to the  Yukawa couplings are
given in the Appendix B.

 For the renormalization of soft SUSY-breaking parameter
$A_q$, we fixed its counterterm by keeping the tree-level relation
of $A_q$, $m_{\tilde{q}_i}$ and $\theta_{\tilde{q}}$ \cite{Aq},
and get the expression as following:
\begin{eqnarray}
&& \delta A_q=\frac{m_{\tilde{q}_1}^2-m_{\tilde{q}_2}^2}{2m_q}
  (2\cos{2\theta_{\tilde{q}}}\delta\theta_{\tilde{q}}
  -\sin{2\theta_{\tilde{q}}}\frac{\delta m_q}{m_q})
  +\frac{\sin{2\theta_{\tilde{q}}}}{2m_q}(\delta m_{\tilde{q}_1}^2
  -\delta m_{\tilde{q}_2}^2) \nonumber \\
  && \hspace{1.3cm}
  +\{\cot\beta, \tan\beta\}\delta \mu
  +\delta\{\cot\beta, \tan\beta\}\mu.
\end{eqnarray}

As for the parameter $\mu$, there are several
schemes\cite{full,mu,muonshell} to fix its counterterm, and here
we use the on-shell renormalization scheme in
Ref.\cite{muonshell}, which gives
\begin{equation}
\delta \mu= \sum_{k=1}^2[m_{\tilde{\chi}_k^+} (\delta U_{k2}V_{k2}
+U_{k2}\delta V_{k2}) +\delta m_{\tilde{\chi}_k^+}U_{k2}V_{k2}],
\end{equation}
where $(U,V)$ are the two $2\times 2$ matrices diagonalizing the
chargino mass matrix, and their counterterms $(\delta U,\delta V)$
are given by
\begin{eqnarray}
\delta U=\frac{1}{4}(\delta Z_R -\delta Z_R^T)U, \\
\delta V=\frac{1}{4}(\delta Z_L -\delta Z_L^T)V.
\end{eqnarray}
The mass shifts $\delta m_{\tilde{\chi}_k^+}$ and the off-diagonal
wave function renormalization constants $\delta Z_{R(L)}$ can be
written as
\begin{eqnarray}
&& \delta m_{\tilde{\chi}_k^+}=\frac{1}{2}{\rm
Re}[m_{\tilde{\chi}_k^+}(\Pi_{kk}^L(m_{\tilde{\chi}_k^+}^2)
+\Pi_{kk}^R(m_{\tilde{\chi}_k^+}^2))
+\Pi_{kk}^{S,L}(m_{\tilde{\chi}_k^+}^2)
+\Pi_{kk}^{S,R}(m_{\tilde{\chi}_k^+}^2)], \\
&& (\delta Z_R)_{ij}=\frac{2}{m_{\tilde{\chi}_i^+}^2
-m_{\tilde{\chi}_j^+}^2}{\rm Re}
[\Pi_{ij}^R(m_{\tilde{\chi}_j^+}^2)m_{\tilde{\chi}_j^+}^2
+\Pi_{ij}^L(m_{\tilde{\chi}_j^+}^2)m_{\tilde{\chi}_i^+}
m_{\tilde{\chi}_j^+} \nonumber \\ && \hspace{6.0cm}
+\Pi_{ij}^{S,R}(m_{\tilde{\chi}_j^+}^2) m_{\tilde{\chi}_i^+}
+\Pi_{ij}^{S,L}(m_{\tilde{\chi}_j^+}^2)m_{\tilde{\chi}_j^+}], \\
&& (\delta Z_L)_{ij}=(\delta Z_R)_{ij} \ \ (L\leftrightarrow R).
\end{eqnarray}
The explicit expressions of the chargino self-energy matrices
$\Pi^{L(R)}$ and $\Pi^{S,L(R)}$ are given in Appendix B.

Finally, the renormalized decay width is then given by
\begin{eqnarray}
\Gamma_s=\Gamma^{(0)}_s +\delta \Gamma^{(v)}_s +\delta
\Gamma^{(c)}_s
\end{eqnarray}
with
\begin{eqnarray}
\delta \Gamma^{(a)}_s =\frac{\lambda^{1/2}(m_{\tilde{t}_2}^2,
m_{\tilde{t}_1}^2, m_{H^0_s}^2)}{8\pi m_{\tilde{t}_2}^3} {\rm Re}
\{M^{(0)\ast}_s \delta M^{(a)}_s\} \ \ \ \ \ (a=v,c).
\end{eqnarray}

\section{Numerical results and conclusion}

We now present some numerical results for the Yukawa
corrections to the decays $\tilde t_2\rightarrow \tilde t_1$+$(h^0,H^0,A^0)$.
The SM input parameters in our calculations were taken
to be $\alpha_{ew}(m_Z)=1/128.8$, $m_W=80.375$GeV,
$m_Z=91.1867$GeV\cite{SM}, $m_t=175.6$GeV, and $m_b=4.25$GeV.

In order to improve the convergence of the perturbation expansion,
using the method presented in Ref.\cite{RAt},   we take into
account the QCD and SUSY QCD running quark masses
$\hat{m}_q(Q)(\hat{m}_t(Q),\hat{m}_b(Q))$ and running trilinear
coupling $\hat{A}_t$  in our calculation(the energy scale Q here
is the mass of the heavier top squark i.e. $m_{\tilde t_2}$).  In
the tree-level $H^0_s\tilde{t}_2 \tilde{t}_1$
 couplings, we use $\hat{m}_t(Q)$ and $\hat{A}_t$ instead of the on-shell parameters.
 While in the calculation of the one-loop corrections, all parameters are on-shell except
 the Yukawa Couplings $h_t,h_b$  taken as the running quark masses.

$\hat{m}_q(Q)$ are evaluated by
the next-to-leading order formula\cite{Runningmb,Runningmt}
\begin{eqnarray} \label{mbQ}
&&\hat{m}_b(Q)=U_6(Q,m_t)U_5(m_t,m_b)m_b(m_b),\nonumber \\
&&\hat{m}_t(Q)=U_6(Q,m_t)m_t(m_t),
\end{eqnarray}
where we have assumed that there are no other colored particles
with masses between scale $Q$ and $m_t$, and
$\hat{m}_b(m_b)=4.25$GeV,$\hat{m}_t(m_t)=175.6$GeV\cite{mb}. The
evolution factor $U_f$ is
\begin{eqnarray}
U_f(Q_2,Q_1)=(\frac{\alpha_s(Q_2)}{\alpha_s(Q_1)})^{d^{(f)}}
[1+\frac{\alpha_s(Q_1)-\alpha_s(Q_2)}{4\pi}J^{(f)}], \nonumber \\
d^{(f)}=\frac{12}{33-2f}, \hspace{1.0cm}
J^{(f)}=-\frac{8982-504f+40f^2}{3(33-2f)^2},
\end{eqnarray}
where $\alpha_s(Q)$ is given by the solutions of the two-loop
renormalization group equations\cite{runningalphas}. When
$Q=400$GeV, the running mass $\hat{m}_b(Q)\sim 2.5$GeV.  In
addition, we also improved the perturbation calculations by the
following replacement \cite{Runningmb,Runningmt}
\begin{eqnarray}\label{replacement}
&&\hat{m}_t(Q) \ \ \rightarrow \ \ \frac{\hat{m}_t(Q)}{1+\Delta
m_t(M_{SUSYQCD})},
 \\
&&\hat{m}_b(Q) \ \ \rightarrow \ \ \frac{\hat{m}_b(Q)}{1+\Delta
m_b(M_{SUSY})},
\end{eqnarray}
where
\begin{eqnarray}\label{gluino}
&& \Delta m_t=-\frac{\alpha_s}{3\pi}\{B_1(0,m_{\tilde
g},m_{\tilde{t}_1}) +B_1(0,m_{\tilde
g},m_{\tilde{t}_2})-\sin2\theta_t(\frac{m_{\tilde g}}{m_t})
[B_0(0,m_{\tilde g},m_{\tilde{t}_1}) -B_0(0,m_{\tilde
g},m_{\tilde{t}_2})]\},
\end{eqnarray}
\begin{eqnarray}\label{gluino}
&& \Delta m_b=\frac{2\alpha_s}{3\pi}M_{\tilde{g}}\mu\tan\beta
I(m_{\tilde{b}_1},m_{\tilde{b}_2},M_{\tilde{g}})
+\frac{h_t^2}{16\pi^2}\mu A_t\tan\beta
I(m_{\tilde{t}_1},m_{\tilde{t}_2},\mu) \nonumber \\
&& \hspace{1.0cm} -\frac{g^2}{16\pi^2}\mu M_2\tan\beta
[\cos^2\theta_{\tilde{t}}I(m_{\tilde{t}_1},M_2,\mu)
+\sin^2\theta_{\tilde{t}}I(m_{\tilde{t}_2},M_2,\mu) \nonumber \\
&& \hspace{4.0cm} +\frac{1}{2}\cos^2\theta_{\tilde{b}}
I(m_{\tilde{b}_1},M_2,\mu) +\frac{1}{2}\sin^2\theta_{\tilde{b}}
I(m_{\tilde{b}_2},M_2,\mu)]
\end{eqnarray}
with
\begin{eqnarray}
I(a,b,c)=\frac{1}{(a^2-b^2)(b^2-c^2)(a^2-c^2)}
(a^2b^2\log\frac{a^2}{b^2} +b^2c^2\log\frac{b^2}{c^2}
+c^2a^2\log\frac{c^2}{a^2}).
\end{eqnarray}

The running trilinear couplings $\hat{A}_t$ can be obtained
according to the procedure of the $\overline{DR}$ renormalization
 where the UV divergence
parameter $\Delta=2/\epsilon-\gamma+\log4\pi$  is set to be zero
\cite{RAt}. First we compute the running stop masses
 $\hat{m}^2_{\tilde t_i}(Q)=\hat{m}^2_{\tilde t_i}+\delta{\hat{m}}^2_{\tilde t_i}$
 and the running mixing angle
 of the top squarks  $\hat{\theta}_{\tilde t}(Q)=\hat{\theta}_{\tilde t}
 +\delta\hat{\theta}_{\tilde t}$,
 where the counterterms $\delta{\hat{m}}^2_{\tilde t_i}$ and $
\delta\hat{\theta}_{\tilde t}$ are given by
\begin{eqnarray}
&&\delta{\hat{m}}^2_{\tilde t_i}=Re[\Sigma^{\tilde
g(g)}_{ii}(m^2_{\tilde t_i}) +\Sigma^{\tilde t(\tilde
g)}_{ii}(m^2_{\tilde t_i})+\Sigma^{\tilde t(\tilde t)}_{ii}],
\nonumber \\
&&\delta\hat{\theta}_{\tilde
t}=\frac{1}{2}\frac{Re\{\Sigma^{\tilde t}_{12}(m^2_{\tilde t_1})
+\Sigma^{\tilde t}_{12}(m^2_{\tilde t_2})\}}{m^2_{\tilde
t_1}-m^2_{\tilde t_2}}.
\end{eqnarray}
Here the explicit expressions of the $\Sigma_{ij}$ functions
arising from the QCD self-energy diagrams are given in
Ref\cite{Runningmt}. Then we can get the running parameter
$\hat{A}_t$ from the formula
\begin{eqnarray}
\hat{m}_t\hat{A}_t=(\hat{m}^2_{\tilde t_1}(Q)-\hat{m}^2_{\tilde
t_2}(Q)) \sin\hat{\theta}_{\tilde t}(Q)\cos\hat{\theta}_{\tilde
t}(Q) +\hat{m}_t\mu\cot\beta.
\end{eqnarray}

The two-loop leading-log relations\cite{Higgs} of the neutral
Higgs boson masses and mixing angles in the MSSM were used. For
$m_{H^+}$ the tree-level formula was used. Other MSSM parameters
were determined as follows:

(i) For the parameters $M_1$, $M_2$ and $\mu$ in the chargino and
neutralino matrix, we take $M_2$ and $\mu$ as the input
parameters, and then use the relation
$M_1=(5/3)(g'^2/g^2)M_2\simeq 0.5M_2$\cite{Haber,M1} to determine
$M_1$. The gluino mass $m_{\tilde{g}}$ was
related to $M_2$ by
$m_{\tilde{g}}=(\alpha_s(m_{\tilde{g}})/\alpha_2)M_2$\cite{Hidaka}.

(ii) For the parameters $m^2_{\tilde{Q},\tilde{U},\tilde{D}}$ and
$A_{t,b}$ in squark mass matrices, we assumed $M_{\tilde
Q}=1.5M_{\tilde U}=1.5M_{\tilde D}$ and $A_t=A_b$ to simplify the
calculations, except for Figs.10-11, where we assumed $M_{\tilde
D}=1.12M_{\tilde Q}$ and $A_t=A_b$ in order to compare with the
SUSY-QCD results in Ref.\cite{QCDcorrection}.

Some typical numerical results of the tree-level decay widths and
the Yukawa corrections are given in Figs.3-12.

Figs.3 - 5 show the  $m_{\tilde{t}_1}$ dependence of the results
of the three decay channels, respectively. Here we take
$m_{A^0}=150$GeV, $\mu=M_2=200$GeV, and $A_t=A_b=600$GeV. The
leading terms of the tree-level amplitudes $M^{(0)}_s$(s=1,2,3)
are given by
\begin{eqnarray}\label{tree}
&& M^{(0)}_1
=\frac{-ig\hat{m}_t}{2m_W\sin\beta}(\hat{A}_t\cos\alpha
+\mu\sin\alpha)\cos2\theta_{\tilde t} \hspace{0.1cm} ,
\end{eqnarray}
\begin{eqnarray}\label{tre}
&&M^{(0)}_2
=\frac{-ig\hat{m}_t}{2m_W\sin\beta}(\hat{A}_t\sin\alpha
-\mu\cos\alpha)\cos2\theta_{\tilde t} \hspace{0.1cm} ,
\end{eqnarray}
\begin{eqnarray}\label{tr}
&&M^{(0)}_3 =\frac{-g\hat{m}_t}{2m_W}(\hat{A}_t\cot\beta+\mu)
\hspace{0.1cm}  .
\end{eqnarray}
For  $m_{\tilde t_1}$=100GeV,
  $\cos\theta_{\tilde{t}}\sim$ (-0.575, -0.574, -0.574) and
 $\cos\alpha\sim$ (0.754, 0.953, 1.000) for
$\tan\beta=$ 4, 10, and 30 , respectively,
 and for
 $m_{\tilde t_1}$=560GeV,
 $\cos\theta_{\tilde{t}}\sim$ (-0.323, -0.332, -0.334) and
 $\cos\alpha\sim$ (0.737, 0.897, 0.992) for $\tan\beta\sim$ 4, 10, and 30, respectively.
In the case of $i=2$,  the two
terms in Eq.(\ref{tre}) have opposite signs, and their magnitudes
get close with the increasing $\tan\beta$ and thus cancel to large
extent for large $\tan\beta$. Therefore, the tree-level decay
widths have the feature of $\Gamma_0(\tan\beta=4)
> \Gamma_0(\tan\beta=10) > \Gamma_0(\tan\beta=30)$ in  most range of the
parameter space, as shown in Fig.4(a). In the case of $i=1$,
 the two terms in Eq.(\ref{tree}) have the same
signs, there are not cancelling effects between them,
so $\Gamma_0$ is larger than the one of the case of $i=2$ for the same values of $\tan\beta$.
In the case of $i=3$, the amplitude contains a term propotional to $\cot\beta$,
so  $\Gamma_0(\tan\beta=4)
> \Gamma_0(\tan\beta=10) > \Gamma_0(\tan\beta=30)$.
From Figs.3-5(b),  one can see that the relative corrections are
sensitive to the value of $\tan\beta$.  For $\tan\beta=$ 4 and 30,
 the magnitudes of the corrections can exceed $30\%$ and $20\%$
, respectively,  for the decay into $h^0$ . For $\tan\beta=$ 10,
the corrections to the widths of the three decay channels are
smaller than ones either in the case of $\tan\beta=$ 4 or in the
case of $\tan\beta=$ 30 .
 In general, for
low $\tan\beta$ the top quark contribution is enhanced while for
high $\tan\beta$ the bottom quark contribution become large, and
for medium $\tan\beta$, there are not any the enhanced effects
from the Yukawa couplings. So the corrections for $\tan\beta=$ 4
or 30 are generally larger than those for $\tan\beta$ =10, as
shown in Figs.3-5(b).  There are some  dips and peaks in
Figs.3-5(b), which arise from the singularities at the threshold
points $m_{\tilde{t}_1}=m_{\tilde{\chi}^0_i}+m_t$ and
$m_{\tilde{t}_2}=m_{\tilde{b}_2}+m_{G^+}(=m_W)$, respectively.

Figs.6-8 give the tree-level decay widths and the Yukawa
corrections as the functions of $m_{A^0}$ for the three decays. We
assumed $m_{\tilde{t}_1}=200$GeV, $\mu=M_2=200$GeV and
$A_t=A_b=1$TeV. The features of the tree level decay widths in
Figs.6-8(a) are similar to Figs.3-5(a) , respectively. From
Figs.6-8(b)
 we can see that the relative corrections decrease or
increase the decay widths depending on $\tan\beta$.
In most  range of the mass of $A^0$, the relative corrections
 vary from 27\% to 33\% for the  decay into $h^0$, -6\% to 20\% for
 the decay into $H^0$, and -9\% to -5\% for the decay into $A^0$.
There are many dips and peaks on the curves in Figs.6-8(b), which
come from the singularities at the threshold points. For example,
at $m_{A^0}=235$GeV in Fig.8(b) , we have the threshold point
$m_{\tilde{b}_1}=m_{\tilde{\chi}^0_4}+m_b$ for $\tan\beta=30$.

In Fig.9 we present the tree-level decay widths and the Yukawa
corrections as the functions of $\mu$ in the case of $\tilde{t}_2
\rightarrow \tilde {t}_1+ H^0_i$ , assuming $\tan\beta=30$,
$m_{\tilde{t}_1}=250$GeV, $M_2=100$GeV, $A_t=250$GeV,
$A_b=-250$GeV and $m_{A^0}=150$GeV. In most of the parameter $\mu$
range , the relative corrections
 are about from 12\% to 32\% for the decay into $h^0$,  and only
 a few percent for the decay into $A^0$ except near
 the zero point of $\Gamma_0$ .
 For the decay into $H^0$,
 when $\mu$ takes certain values (near about -26 GeV),  $\Gamma_0$ gets very small
 ($<10^{-4}$GeV),  and the relative corrections near these values do not have
 a physical meaning. So we cut off the corrections, since perturbation
 theory breaks down here. In order to improve the results, we use the running
 higgsino mass parameter $\hat\mu(Q)=\mu+\delta\hat\mu(Q)$ in the tree-level
 coupling, and find that
 the convergence of the perturbation expansion becomes  better as shown by the dashed
 line in Fig.9(b),
 where  the region of the parameter $\mu$
  of breaking down the perturbation theory gets smaller
 (Note that, in fact,  the parameter range
  $|\mu|<180$ GeV has been excluded by phenomenology at LEP and Tevatron
  \cite{RAt,excludemu} ).
There are many dips and peaks on the curves in Fig.9(b), which
come from the singularities at the threshold points. For example,
at $\mu=-216$ GeV on the solid curve in Fig.9(b) , we have the
threshold point $m_{\tilde{t}_2}=m_{\chi^0_4}+m_t$ for the decay
into $H^0$.

In Figs.10-11 we compare the results with the ones presented in an
earlier literature \cite{QCDC} where  the ${\cal O}(\alpha_s)$
SUSY-QCD corrections to the same three decay processes have been
calculated. We present the tree-level decay widths and the Yukawa
corrected decay widths as the functions of $m_{\tilde{t}_2}$ and
$m_{A^0}$ in Figs.10 and 11, respectively . For comparation, we
take the same input parameters as in the Ref.\cite{QCDC} :
$\tan\beta=3$, $\cos\theta_{\tilde t}=0.26$,
$m_{\tilde{t}_1}=250$GeV, $m_{\tilde g}=600$GeV, $\mu=550$GeV  in
Figs.10-11,  and $m_{A^0}=150$GeV in Fig.10,
$m_{\tilde{t}_2}=600$GeV  in Fig.11.  In both Figs, we assumed
$M_{\tilde D}=1.12M_{\tilde Q}$. Our numerical results of the tree
level decay widths agree with their results except a little
difference, which is due to the running effects were used in our
calculation but not in Ref.\cite{QCDC}. The relative corrections
in Fig.10 vary from -22\% to 26\% for the decay into $h^0$,  -60\%
to -4\% for the decay into $H^0$, and -5\% to 0\% for the decay
into $A^0$.  The relative corrections in Fig.11 vary from -1\% to
23\% for the decay into $h^0$, -24\% to 60\% for the decay into
$H^0$, -4\% to -1\% for the decay into $A^0$. After comparing with
Figs.3 and 5 in Ref.\cite{QCDC}, we can see that the Yukawa
corrections are comparable to the ${\cal O}(\alpha_s)$ SUSY-QCD
corrections for the decays into $h^0$ and $H^0$, but smaller than
the ${\cal O}(\alpha_s)$ SUSY-QCD corrections for the decays into
$A^0$. There are two dips at $m_{A^0}=348$GeV and 352GeV on the
solid curve of the decay into $h_0$ in Fig.11, which
 come from the singularities at the threshold points
$m_{\tilde{t}_2}=m_{\tilde{t}_1}+m_{H^0}.$

Finally, in Fig.12 we show the numerical improvement of the Yukawa
corrections as a function of $\tan\beta$ in five ways of
perturbative expansion: (i) the strict on-shell scheme (the dotted
line), where the top quark pole mass 175.6GeV, the bottom quark
pole mass 4.25GeV,  the on-shell trilinear coupling $A_t$ and the
 higssino mass parameter $\mu$
 were used, (ii) the improved scheme (the
solid line), in which the QCD, SUSY-QCD, and SUSY-Electroweak
running quark masses $\hat{m}_q(Q)$ and the running trilinear
coupling $\hat{A}_t(Q)$ were used, (iii) the complete improved
scheme (the dashed line),
 in which the SUSY-Electroweak running parameter $\mu$  was also used as well
as the same running parameters as in (ii),
  \hspace{0.3cm}(iv) the  $\hat{m}_t(Q)$ running scheme (the
dash-dotted line), in which  only  the running top quark mass was
used,
  and (v) the  $\hat{m}_b(Q)$ running scheme (the
dash-dot-dotted line), in which  only  the  running bottom quark
mass was used.
 Here we assumed $m_{\tilde{t}_1}=250$GeV,
$M_2=200$GeV, $A_t=A_b=900$GeV, $\mu=200$GeV, $m_{A^0}=150$GeV and
$M_{\tilde Q}=1.5M_{\tilde U}=1.5M_{\tilde D}$. One can see that,
 the effect of the running
of the top quark mass on the corrections can not be neglected
 for low $\tan\beta(<10)$,  while the effect of the
  running of the bottom quark mass is quite significant for large $\tan\beta(>40)$ .
  The whole running effects with
or without the running of the parameter $\mu$  both  make the
convergence of the perturbation expansion much better. The
relative corrections approach smoothly -5.0\% and 14.3\%  with the
increasing $\tan\beta$ for the improved scheme and complete
improved scheme, as shown by  the solid line and the dashed line
in Fig.12,  respectively.

In conclusion, we have calculated the Yukawa corrections to the
widths of the heavier top squark decays into lighter top squarks
and neutral Higgs bosons in the MSSM. These corrections depend on
the masses of the neutral Higgs bosons and the lighter or heavier
top squark, and the parameter $\mu$. For favorable parameter
values, the corrections decrease or increase the tree-level decay
widths significantly. Especially, for high values of
$\tan\beta$(=30) or low values of $\tan\beta$(=4), the magnitudes
of the corrections exceed at least $20\%$ for the  decay into
$h^0$ and $H^0$, which are comparable to the  ${\cal O}(\alpha_s)$
SUSY-QCD corrections. But for the decay into $A^0$,
 the corrections are smaller and the magnitudes of them
are less than $10\%$ in most of the parameter space. The numerical
calculations also show that using the running quark masses and the
running trilinear coupling $A_t$ , which include the QCD,
SUSY-QCD, and SUSY-Electroweak effects and resume all high order
($\tan\beta$)-enhanced effects, can vastly improve the convergence
of the perturbation expansion. We also discuss the effects of the
running of the higgsino mass parameter $\mu$ on the corrections,
and find that they are significant, too, especially for large
$\tan\beta$.
\section*{Acknowledgements}

This work was supported in part by the National Natural Science
Foundation of China, the Doctoral Program Foundation of Higher
Education of China.

\section*{Appendix A}

The following couplings are given in order $O(h_t, h_b)$.

1. squark -- squark -- Higgs boson

(a) squark -- squark -- $h^0$
\begin{eqnarray}
\hat{G}^{\tilde{q}}_1= \left(\begin{array}{cc} -\sqrt{2}m_qh_q
\left\{\begin{array}{c}c_\alpha \\ -s_\alpha\end{array}\right\}
& -\frac{1}{\sqrt{2}}h_q(A_q \left\{\begin{array}{c}c_\alpha \\
-s_\alpha\end{array}\right\} +\mu \left \{\begin{array}{c}
s_\alpha \\ -c_\alpha \end{array} \right\}) \\
-\frac{1}{\sqrt{2}}h_q(A_q \left\{\begin{array}{c} c_\alpha \\
-s_\alpha \end{array} \right\} +\mu \left \{\begin{array}{c}
s_\alpha \\ -c_\alpha \end{array} \right\}) & -\sqrt{2}m_qh_q
\left\{\begin{array}{c} c_\alpha \\ -s_\alpha \end{array} \right\}
\end{array} \right)
\end{eqnarray}
for $\left\{\begin{array}{c} {\rm up} \\ {\rm down} \end{array}
\right\}$ type squarks, respectively. We use the abbreviations
$s_\alpha=\sin\alpha$, $c_\alpha=\cos\alpha$. $\alpha$ is the
mixing angle in the CP even neutral Higgs boson sector.

(b) squark -- squark -- $H^0$
\begin{eqnarray}
\hat{G}^{\tilde{q}}_2= \left(\begin{array}{cc} -\sqrt{2}m_qh_q
\left\{\begin{array}{c}s_\alpha \\ c_\alpha\end{array}\right\}
& -\frac{1}{\sqrt{2}}h_q(A_q \left\{\begin{array}{c} s_\alpha \\
c_\alpha\end{array}\right\} -\mu \left \{\begin{array}{c}
c_\alpha \\ s_\alpha \end{array} \right\}) \\
-\frac{1}{\sqrt{2}}h_q(A_q \left\{\begin{array}{c} s_\alpha \\
c_\alpha \end{array} \right\} -\mu \left \{\begin{array}{c}
c_\alpha \\ s_\alpha \end{array} \right\}) & -\sqrt{2}m_qh_q
\left\{\begin{array}{c} s_\alpha \\ c_\alpha \end{array} \right\}
\end{array} \right)
\end{eqnarray}

(c) squark -- squark -- $A^0$
\begin{eqnarray}
\hat{G}^{\tilde{q}}_3=i\frac{gm_q}{2m_W} \left(\begin{array}{cc} 0
& -A_q\left\{\begin{array}{c}\cot\beta \\ \tan\beta
\end{array}\right\} -\mu \\ A_q\left\{\begin{array}{c}\cot\beta
\\ \tan\beta \end{array}\right\} +\mu & 0 \end{array} \right)
\end{eqnarray}

(d) squark -- squark -- $G^0$
\begin{eqnarray}
\hat{G}^{\tilde{q}}_4=-i\frac{gm_q}{2m_W} \left(\begin{array}{cc}
0 & -A_q +\mu\left\{\begin{array}{c}\cot\beta \\ \tan\beta
\end{array}\right\} \\ A_q -\mu\left\{\begin{array}{c}\cot\beta
\\ \tan\beta \end{array}\right\} & 0 \end{array} \right)
\end{eqnarray}

(e) squark -- squark -- $H^\pm$
\begin{eqnarray}
\hat{G}^{\tilde{b}}_5=(\hat{G}^{\tilde{t}}_5)^T =
\frac{g}{\sqrt{2}m_W}\left(\begin{array}{cc}
m_b^2\tan\beta +m_t^2\cot\beta & m_t(A_t\cot\beta +\mu) \\
m_b(A_b\tan\beta +\mu) & 2m_tm_b/\sin2\beta
\end{array} \right)
\end{eqnarray}

(f) squark -- squark -- $G^\pm$
\begin{eqnarray}
\hat{G}^{\tilde{b}}_6=(\hat{G}^{\tilde{t}}_6)^T =
\frac{-g}{\sqrt{2}m_W}\left(\begin{array}{cc}
m_t^2 -m_b^2 & m_t(A_t -\mu\cot\beta) \\
m_b(\mu\tan\beta -A_b) & 0
\end{array} \right)
\end{eqnarray}

2. quark -- quark -- Higgs boson
\begin{eqnarray}
&& a_k^q=(\frac{1}{\sqrt{2}}h_q\left\{\begin{array}{c} -c_\alpha
\\ s_\alpha \end{array} \right\}, -\frac{1}{\sqrt{2}}h_q
\left\{\begin{array}{c} s_\alpha \\ c_\alpha \end{array} \right\},
-\frac{i}{\sqrt{2}}h_q\left\{\begin{array}{c} \cos\beta
\\ \sin\beta \end{array} \right\},
\frac{-ig}{2m_W}\left\{\begin{array}{c} -m_t \\ m_b
\end{array}\right\}, \nonumber \\
&& \hspace{3.0cm} \left\{\begin{array}{c} h_b\sin\beta
\\ h_t\cos\beta \end{array} \right\},
\frac{g}{\sqrt{2}m_W}\left\{\begin{array}{c} -m_b \\m_t
\end{array}\right\})
\end{eqnarray}

\begin{eqnarray}
&& b_k^q=(\frac{1}{\sqrt{2}}h_q\left\{\begin{array}{c} -c_\alpha
\\ s_\alpha \end{array} \right\}, -\frac{1}{\sqrt{2}}h_q
\left\{\begin{array}{c} s_\alpha \\ c_\alpha \end{array} \right\},
-\frac{i}{\sqrt{2}}h_q\left\{\begin{array}{c} \cos\beta
\\ \sin\beta \end{array} \right\},
\frac{-ig}{2m_W}\left\{\begin{array}{c} m_t \\ -m_b
\end{array}\right\}, \nonumber \\
&& \hspace{3.0cm} h_q \left\{\begin{array}{c} \cos\beta
\\ \sin\beta \end{array} \right\},
\frac{g}{\sqrt{2}m_W}\left\{\begin{array}{c} m_t \\ -m_b
\end{array}\right\})
\end{eqnarray}

3. quark -- squark -- neutralino
\begin{eqnarray}
a_{ik}^{\tilde{q}}=-R_{i2}^{\tilde{q}}Y_q \left\{\begin{array}{c}
N_{k4} \\ N_{k3} \end{array} \right\}, \hspace{1.5cm}
b_{ik}^{\tilde{q}}=-R_{i1}^{\tilde{q}}Y_q \left\{\begin{array}{c}
N_{k4}^\ast \\ N_{k3}^\ast \end{array} \right\}
\end{eqnarray}
Here $N$ is the $4\times 4$ unitary matrix diagonalizing the
neutral gaugino-higgsino mass matrix \cite{Haber,M1}.

4. quark -- squark -- chargino
\begin{eqnarray}
l_{ik}^{\tilde{q}}=R_{i2}^{\tilde{q}}Y_q \left\{\begin{array}{c}
V_{k2} \\ U_{k2} \end{array} \right\}, \hspace{1.5cm}
k_{ik}^{\tilde{q}}=R_{i1}^{\tilde{q}} \left\{\begin{array}{c}
Y_bU_{k2} \\ Y_tV_{k2} \end{array} \right\}.
\end{eqnarray}
Here $U$ and $V$ are the $2\times 2$ unitary matrices
diagonalizing the charged gaugino--higgsino mass matrix
\cite{Haber,M1}.

5. squark -- squark -- Higgs boson -- Higgs boson

(a) squark -- squark -- $H^-$ -- $H_k$ (k=1,2)
\begin{eqnarray}
\hat{G}_{5k}^{\tilde{b}}=(\hat{G}_{5k}^{\tilde{t}})^T
=\frac{g^2}{2\sqrt{2}m_W^2}\left(\begin{array}{cc} m_t^2S_k
+m_b^2T_k & 0 \\ 0 & \frac{2m_tm_b}{\sin2\beta}  V_k
\end{array}\right)
\end{eqnarray}
with
\begin{eqnarray}
&& S_k=(\cos\alpha \cos\beta/\sin^2\beta, \ \ \  \sin\alpha
\cos\beta/\sin^2\beta) \\ && T_k=(-\sin\alpha
\sin\beta/\cos^2\beta, \ \ \  \cos\alpha
\sin\beta/\cos^2\beta) \\
&& V_k=(\sin(\beta -\alpha), \ \ \ \cos(\beta -\alpha))
\end{eqnarray}

(b) squark -- squark -- $H^-$ -- $H^+$
\begin{eqnarray}
\hat{G}_{55}^{\tilde{q}}=\left(\begin{array}{cc}
-\left\{\begin{array}{c}h_b^2\sin^2\beta \\ h_t^2\cos^2\beta
\end{array}\right\} & 0 \\ 0 & -h_q^2\left\{\begin{array}{c}
\cos^2\beta \\ \sin^2\beta \end{array}\right\}
\end{array}\right)
\end{eqnarray}

(c) squark -- squark -- $H^-$ -- $G^+$
\begin{eqnarray}
\hat{G}_{56}^{\tilde{q}}=-\frac{g^2}{2m_W^2}\left(\begin{array}{cc}
\left\{\begin{array}{c}-m_b^2\tan\beta \\ m_t^2\cot\beta
\end{array} \right\} & 0 \\ 0 & m_q^2 \left\{\begin{array}{c}
\cot\beta \\ -\tan\beta \end{array} \right\}\end{array} \right)
\end{eqnarray}

(d) squark -- squark -- $G^-$ -- $H_k$ (k=1,2,3)
\begin{eqnarray}
\hat{G}_{6k}^{\tilde{b}}=(\hat{G}_{6k}^{\tilde{t}})^T
=\frac{g^2}{2\sqrt{2}m_W^2}\left(\begin{array}{cc}
m_t^2\hspace{0.2cm} SG_k +m_b^2\hspace{0.2cm}  TG_k & 0 \\ 0 &
2m_tm_b/\sin2\beta \hspace{0.2cm}VG_k
\end{array}\right)
\end{eqnarray}
with
\begin{eqnarray}
&& SG_k=(\cos\alpha /\sin\beta, \ \ \  \sin\alpha /\sin\beta \ \ \
i\cot\beta )
\\ && TG_k=(\sin\alpha /\cos\beta, \ \ \  -\cos\alpha
/\cos\beta \ \ \ i\tan\beta) \\
&& VG_k=(-\cos(\beta -\alpha), \ \ \ \sin(\beta -\alpha) \ \ \ -i)
\end{eqnarray}

(e) squark -- squark -- $H_k$ -- $H_k$ (k=1,2,3)
\begin{eqnarray}
\hat{G}_{kk}^{\tilde{t}} =\left(\begin{array}{cc}
\frac{-g^2}{2m_W^2}m_t^2\hspace{0.1cm}{D1}_k& 0 \\ 0
&\frac{-g^2}{2m_W^2}m_t^2\hspace{0.1cm}{D1}_k
\end{array}\right)
\end{eqnarray}
with
\begin{eqnarray}
 {D1}_k=(\sin^2\alpha /\sin^2\beta, \ \ \  \cos^2\alpha /\sin^2\beta,
\ \ \ \cot^2\beta )
\end{eqnarray}

\begin{eqnarray}
\hat{G}_{kk}^{\tilde{b}} =\left(\begin{array}{cc}
\frac{-g^2}{2m_W^2}m_b^2\hspace{0.1cm}{D2}_k& 0 \\ 0
&\frac{-g^2}{2m_W^2}m_b^2\hspace{0.1cm}{D2}_k
\end{array}\right)
\end{eqnarray}
with
\begin{eqnarray}
 {D2}_k=(\cos^2\alpha /\sin^2\beta, \ \ \  \sin^2\alpha /\sin^2\beta,
\ \ \ \tan^2\beta )
\end{eqnarray}

(f) squark -- squark -- $H^0$ -- $h^0$
\begin{eqnarray}
\hat{G}_{12}^{\tilde{b}} =\left(\begin{array}{cc}
\frac{-g^2m_b^2\sin2\alpha}{4m_W^2}\hspace{0.1cm}D2& 0 \\ 0
&\frac{-g^2m_b^2\sin2\alpha}{4m_W^2}\hspace{0.1cm}D2
\end{array}\right)
\end{eqnarray}
with $D2=-1/\cos^2\beta$

\begin{eqnarray}
\hat{G}_{12}^{\tilde{t}} =\left(\begin{array}{cc}
\frac{-g^2m_t^2\sin2\alpha}{4m_W^2}\hspace{0.1cm}D1& 0 \\ 0
&\frac{-g^2m_t^2\sin2\alpha}{4m_W^2}\hspace{0.1cm}D1
\end{array}\right)
\end{eqnarray}
with $D1=-1/\sin^2\beta$

(g) squark -- squark -- $A^0$ -- $G^0$
\begin{eqnarray}
\hat{G}_{35}^{\tilde{b}} =\left(\begin{array}{cc}
\frac{-g^2m_b^2\sin2\beta}{4m_W^2}\hspace{0.1cm}D2& 0 \\ 0
&\frac{-g^2m_b^2\sin2\beta}{4m_W^2}\hspace{0.1cm}D2
\end{array}\right)
\end{eqnarray}

\begin{eqnarray}
\hat{G}_{35}^{\tilde{t}} =\left(\begin{array}{cc}
\frac{-g^2m_b^2\sin2\beta}{4m_W^2}\hspace{0.1cm}D1& 0 \\ 0
&\frac{-g^2m_b^2\sin2\beta}{4m_W^2}\hspace{0.1cm}D1
\end{array}\right)
\end{eqnarray}
\\
Finally, we
define$\hat{G}_{ji}^{\tilde{q}}=\hat{G}_{ij}^{\tilde{q}},$ and
also $\hat{G}_{3(4) k}^{\tilde{q}}=0$, when k=1,2,5, i.e. there
are no $A^0(G^0)h^0\tilde q\tilde q$,$A^0(G^0)H^0\tilde q\tilde
q$, or $A^0(G^0)H^+\tilde q\tilde q$  couplings.
\section*{Appendix B }
We define $q=t$ and $b$, $q'$ the $SU(2)_L$ partner of $q$, and
$q''=q$ for $k=1...4$ and $q''=q'$ for $k=5,6$. Then we have
\begin{eqnarray}
&& \frac{\delta m_W^2}{m_W^2}=\frac{g^2}{
  16\pi^2m_W^2}[m_b^2 +m_t^2 -A_0(m_t^2) -A_0(m_b^2) -m_t^2B_0 -(m_t^2
  -m_b^2)B_1] \nonumber \\
  && \hspace{1.3cm}
  (m_W^2,m_b,m_t), \nonumber \\
&& \frac{\delta m_Z^2}{m_Z^2}=\frac{3g^2}{8\pi^2m_W^2}
  \sum_{q=t,b}\{\frac{1}{3}[(I_{3L}^q -e_q\sin^2\theta_W)^2
  +e_q^2\sin^4\theta_W][2m_q^2 -2A_0(m_q^2) -m_q^2B_0]  \nonumber \\
  && \hspace{1.3cm}
  -2m_q^2e_q\sin^2\theta_W(I_{3L}^q -e_q\sin^2\theta_W)B_0\}
  (m_Z^2,m_q,m_q), \nonumber \\
&& \delta Z_{H^-}=\frac{3g^2}{16\pi^2}[(m_t^2\cot^2\beta+m_b^2\tan^2\beta)
(m_{H^+}^2G_1 +B_1 -m_t^2G_0)
  -2m_t^2m_b^2G_0] \nonumber \\
  && \hspace{1.4cm}
  (m_{H^+}^2,m_t,m_b)
  +\frac{3}{16\pi^2}\sum_{i,j}(G_5^{\tilde{t}})_{ij}(G_5^{\tilde{t}})_{ij}
  G_0(m_{H^+}^2,m_{\tilde{t}_i},m_{\tilde{b}_j}), \nonumber \\
&& \delta
Z_{h^0}=\frac{3g^2m_t^2\cos^2\alpha}{16\pi^2m_W^2\sin^2\beta}
(-2m_t^2G_0+B_1+m_h^2G_1)(m_{h_0}^2,m_t,m_t) \nonumber \\
  && \hspace{1.4cm}
  +\frac{3g^2m_b^2\sin^2\alpha}{16\pi^2m_W^2\cos^2\beta}
(-2m_b^2G_0+B_1+m_h^2G_1)(m_{h_0}^2,m_b,m_b) \nonumber \\
  && \hspace{1.4cm}
  +\frac{3}{16\pi^2}\sum_{i,j}(G_1^{\tilde{t}})_{ij}(G_1^{\tilde{t}})_{ij}
  G_0(m_{h_0}^2,m_{\tilde{t}_i},m_{\tilde{b}_j}), \nonumber \\
  && \delta Z_{H^0}=\frac{3g^2m_t^2\sin^2\alpha}{16\pi^2m_W^2\sin^2\beta}
(-2m_t^2G_0+B_1+m_H^2G_1)(m_{H_0}^2,m_t,m_t) \nonumber \\
  && \hspace{1.4cm}
  +\frac{3g^2m_b^2\cos^2\alpha}{16\pi^2m_W^2\cos^2\beta}
(-2m_b^2G_0+B_1+m_H^2G_1)(m_{H_0}^2,m_b,m_b) \nonumber \\
  && \hspace{1.4cm}
  +\frac{3}{16\pi^2}\sum_{i,j}(G_2^{\tilde{t}})_{ij}(G_2^{\tilde{t}})_{ij}
  G_0(m_{H_0}^2,m_{\tilde{t}_i},m_{\tilde{b}_j}), \nonumber \\
&& \delta
Z_{A^0}=\frac{3g^2m_t^2\cos^2\alpha}{16\pi^2m_W^2\sin^2\beta}
(-2m_t^2G_0+B_1+m_h^2G_1) (m_{A_0}^2,m_t,m_t)\nonumber \\
  && \hspace{1.4cm}
  +\frac{3g^2m_b^2\sin^2\alpha}{16\pi^2m_W^2\cos^2\beta}
(-2m_b^2G_0+B_1+m_h^2G_1)(m_{A_0}^2,m_b,m_b) \nonumber \\
  && \hspace{1.4cm}
  +\frac{3}{16\pi^2}\sum_{i,j}(G_3^{\tilde{t}})_{ij}(G_3^{\tilde{t}})_{ij}
  G_0(m_{A_0}^2,m_{\tilde{t}_i},m_{\tilde{b}_j}), \nonumber \\
&& T_{H_k}=\frac{-3gm_t^2}{8\pi^2m_W\sin\beta}A_0(m_t^2)
+\frac{-3gm_b^2}{8\pi^2m_W\cos\beta}A_0(m_b^2)
 -\sum_{q= t,b}\sum_j(G_k^{\tilde{q}})_{jj}A_0(m_{\tilde
  {q}_j}^2)\}, \nonumber \\
&&
\Sigma_{GH}=-\frac{3g^2}{16\pi^2m_W^2}(m_t^2\cot\beta-m_b^2\tan\beta)
(m_t^2B_0+A_0(m_b^2)+m_{H^+}^2B_1)
   +m_t^2m_b^2(\tan\beta
  \nonumber \\
  && \hspace{1.3cm}
-\cot\beta)B_0(m_{H^+}^2,m_t,m_b)
  +\frac{-3}{16\pi^2}\sum_{j,l}
  (G_5^{\tilde{t}})_{jl}(G_6^{\tilde{t}})_{lj}B_0
  (m_{H^+}^2,m_{\tilde{t}_l},m_{\tilde{b}_j}) \nonumber \\
  && \hspace{1.3cm}
  +\frac{3}{16\pi^2}\sum_{q=t,b}\sum_ji(G_{56}^{\tilde{q}})_{jj}A_0(m_{\tilde{q}_j}^2),
  \nonumber \\
&& \frac{\delta m_t}{m_t}=\frac{1}{16\pi^2}
  \sum_{k=1}^6[\frac{m_{t''}}{m_t} a_k^ta_k^{t''}B_0
  -\frac{1}{2}(a_k^tb_k^{t''}
  +b_k^ta_k^{t''})B_1](m_t^2,m_{t''},m_{H_k}) \nonumber \\
  && \hspace{1.3cm}
  +\frac{g^2}{16\pi^2}\sum_{k=1}^4\sum_j[\frac{m_{\tilde{\chi}_k^0}}{m_t}
  a_{jk}^{\tilde{t}}b_{jk}^{\tilde{t}\ast}B_0
  +\frac{1}{2}(|a_{jk}^{\tilde{t}}|^2
  +|b_{jk}^{\tilde{t}}|^2)(B_0+B_1)](m_t^2,m_{\tilde{t}_j},m_{\tilde{\chi}_k^0})
  \nonumber \\
  && \hspace{1.3cm}
  +\frac{g^2}{16\pi^2}\sum_{k=1}^2\sum_j[\frac{m_{\tilde{\chi}_k^+}}
  {m_b}l_{jk}^{\tilde{b}}k_{jk}^{\tilde{b}}B_0
  +\frac{1}{2}(|l_{jk}^{\tilde{b}}|^2
  +|k_{jk}^{\tilde{b}}|^2)
  (B_0+B_1)](m_t^2,m_{\tilde{b}_j}},m_{\tilde{\chi}_k^+),
  \nonumber \\
&& \delta m_{\tilde{t}_i}^2=\frac{1}{16\pi^2}\{\sum_{k=1}^6\sum_j
  (G_k^{\tilde{t}})_{ij}(G_k^{\tilde{t}''})_{ji}B_0
  (m_{\tilde{t}_i}^2,m_{\tilde{t}_j''},m_{H_k})
  -2g^2\sum_{k=1}^4[(|a_{ik}^{\tilde{t}}|^2 +|b_{ik}^{\tilde{t}}|^2)
  \nonumber \\
  && \hspace{1.3cm}
  \times (m_{\tilde{t}_i}^2B_1
  +A_0(m_{\tilde{\chi}_k^0}^2) +m_t^2B_0)
  +2m_tm_{\tilde{\chi}_k^0}{\rm Re}(a_{ik}^{\tilde{t}}b_{ik}^{\tilde{t}\ast})
  B_0](m_{\tilde{t}_i}^2,m_t,m_{\tilde{\chi}_k^0}) \nonumber \\
  && \hspace{1.3cm}
  -2g^2\sum_{k=1}^2[(|l_{ik}^{\tilde{t}}|^2
  +|k_{ik}^{\tilde{t}}|^2)(m_{\tilde{t}_i'}^2B_1
  +A_0(m_{\tilde{\chi}_k^+}^2) +m_{t'}^2B_0) \nonumber \\
  && \hspace{1.3cm}
  +2m_{t'}m_{\tilde{\chi}_k^+}{\rm Re}(l_{ik}^{\tilde{t}}k_{ik}^{\tilde{t}\ast})
  B_0](m_{\tilde{t}_i}^2,m_{t'},m_{\tilde{\chi}_k^+})\}, \nonumber \\
&& \delta Z_{\tilde{t}_i}=\frac{1}{16\pi^2}\{\sum_{k=1}^6\sum_j
  (G_k^{\tilde{t}})_{ij}(G_k^{\tilde{t}''})_{ji}G_0
  (m_{\tilde{t}_i}^2,m_{\tilde{t}_j''},m_{H_k})
  +2g^2\sum_{k=1}^4[(|a_{ik}^{\tilde{t}}|^2 +|b_{ik}^{\tilde{t}}|^2)
  \nonumber \\
  && \hspace{1.3cm}
  \times (B_1 +m_{\tilde{t}_i}^2G_1 -m_t^2G_0)
  -2m_tm_{\tilde{\chi}_k^0}{\rm Re}(a_{ik}^{\tilde{t}}b_{ik}^{\tilde{t}\ast})
  G_0](m_{\tilde{t}_i}^2,m_t,m_{\tilde{\chi}_k^0}) \nonumber \\
  && \hspace{1.3cm}
  +2g^2\sum_{k=1}^2[(|l_{ik}^{\tilde{t}}|^2
  +|k_{ik}^{\tilde{t}}|^2)(B_1 +m_{\tilde{t}_i'}^2G_1
  -m_{t'}^2G_0) \nonumber \\
  && \hspace{1.3cm}
  -2m_{q'}m_{\tilde{\chi}_k^+}{\rm Re}(l_{ik}^{\tilde{q}}k_{ik}^{\tilde{q}\ast})
  G_0](m_{\tilde{q}_i}^2,m_{q'},m_{\tilde{\chi}_k^+})\},
  \nonumber \\
&&
\Sigma_{Hh}(p^2)=\frac{-3g^2m_t^2\sin2\alpha}{32\pi^2m_W^2\sin^2\beta}
[(2m_t^2B_0+p^2B_1)(p^2,m_t,m_t)+A_0(m_t^2)] \nonumber \\
&&\hspace{1.3cm}+\frac{3g^2m_b^2\sin2\alpha}{32\pi^2m_W^2\cos^2\beta}
[(2m_b^2B_0+p^2B_1)(p^2,m_b,m_b)+A_0(m_b^2)] \nonumber \\
&& \hspace{1.3cm}+\frac{3}{16\pi^2}
\sum_q\sum_{i,j}(G_1^{\tilde{q}})_{ji}(G_2^{\tilde{q}})_{ij}
B_0(p^2,m_{\tilde{q}_j},m_{\tilde{q}_i}) \nonumber
\\
&& \hspace{1.3cm}+\frac{3i}{16\pi^2}
\sum_q\sum_i (G_{12}^{\tilde{q}})_{ii}A_0(m_{\tilde{q}_i}) \nonumber
\\
&&\delta Z_{H^0h^0}=\frac{\Sigma_{Hh}(h^2_0)}{m^2_{H^0}-m^2_{h^0}},\hspace{0.6cm}
\delta Z_{h^0H^0}=\frac{\Sigma_{Hh}(H^2_0)}{m^2_{h^0}-m^2_{H^0}}\nonumber
\\
&& \Sigma_{12}^{\tilde{t}}(p^2)=\frac{1}{16\pi^2}
  \{\sum_{k=1}^6\sum_j
  (G_k^{\tilde{t}})_{1j}(G_k^{\tilde{t}''})_{j2}B_0
  (p^2,m_{\tilde{t}_j''},m_{H_k})
  -2g^2\sum_{k=1}^4[(a_{1k}^{\tilde{t}}a_{2k}^{\tilde{t}\ast}
  +b_{1k}^{\tilde{t}}b_{2k}^{\tilde{t}\ast})  \nonumber \\
  && \hspace{1.3cm}
  \times (p^2B_1 +A_0(m_{\tilde{\chi}_k^0}^2) +m_t^2B_0)
  +m_tm_{\tilde{\chi}_k^0}(a_{1k}^{\tilde{t}}b_{2k}^{\tilde{t}\ast}
  +a_{2k}^{\tilde{t}\ast}b_{1k}^{\tilde{t}})
  B_0](p^2,m_t,m_{\tilde{\chi}_k^0}) \nonumber \\
  && \hspace{1.3cm}
  -2g^2\sum_{k=1}^2[(l_{1k}^{\tilde{t}}l_{2k}^{\tilde{t}\ast}
  +k_{1k}^{\tilde{t}}k_{2k}^{\tilde{t}\ast})(p^2B_1
  +A_0(m_{\tilde{\chi}_k^+}^2) +m_{t'}^2B_0) \nonumber \\
  && \hspace{1.3cm}
  +m_{t'}m_{\tilde{\chi}_k^+}(l_{1k}^{\tilde{t}}k_{2k}^{\tilde{t}\ast}
  +l_{2k}^{\tilde{t}\ast}k_{1k}^{\tilde{t}})
  B_0](p^2,m_{t'},m_{\tilde{\chi}_k^+})\}, \nonumber \\
&& \delta \theta_{\tilde{t}} +\delta Z_{21}^{\tilde{t}}
  =\frac{1}{2(m_{\tilde{t}_1}^2
  -m_{\tilde{t}_2}^2)}[\Sigma_{12}^{\tilde{t}}(m_{\tilde{t}_2}^2)
  -\Sigma_{12}^{\tilde{t}}(m_{\tilde{t}_1}^2)], \nonumber \\
&& \Pi_{ij}^L(p^2)=-\frac{3}{16\pi^2}\sum_{k=1}^2
[l_{ki}^{\tilde{t}} l_{kj}^{\tilde{t}}B_1(p^2,m_b,m_{\tilde{t}_k})
+k_{ki}^{\tilde{b}}
k_{kj}^{\tilde{b}}B_1(p^2,m_t,m_{\tilde{b}_k})], \nonumber \\
&& \Pi_{ij}^R(p^2)=-\frac{3}{16\pi^2}\sum_{k=1}^2
[k_{ki}^{\tilde{t}} k_{kj}^{\tilde{t}}B_1(p^2,m_b,m_{\tilde{t}_k})
+l_{ki}^{\tilde{b}}
l_{kj}^{\tilde{b}}B_1(p^2,m_t,m_{\tilde{b}_k})], \nonumber \\
&& \Pi_{ij}^{S,L}(p^2)=\frac{3}{16\pi^2}\sum_{k=1}^2
[m_bk_{ki}^{\tilde{t}}
l_{kj}^{\tilde{t}}B_0(p^2,m_b,m_{\tilde{t}_k})
+m_tl_{ki}^{\tilde{b}}
k_{kj}^{\tilde{b}}B_0(p^2,m_t,m_{\tilde{b}_k})], \nonumber \\
&& \Pi_{ij}^{S,R}(p^2)=\frac{3}{16\pi^2}\sum_{k=1}^2
[m_bl_{ki}^{\tilde{t}}
k_{kj}^{\tilde{t}}B_0(p^2,m_b,m_{\tilde{t}_k})
+m_tk_{ki}^{\tilde{b}}
l_{kj}^{\tilde{b}}B_0(p^2,m_t,m_{\tilde{b}_k})]. \nonumber
\end{eqnarray}
Here $A_0$ and $B_1$ are one- and two-point Feynman
integrals\cite{ABCD}, respectively, and $G_1=\partial B_1/\partial
p^2$,$G_0=-\partial B_0/\partial
p^2$.


\newpage
\begin{figure}
\begin{center}
\begin{picture}(110,110)(0,0)
\DashLine(20,60)(55,60){3} \DashLine(55,60)(85,90){3}
\DashLine(55,60)(85,30){3} \Vertex(55,60){1}
\Text(15,66)[]{$\tilde{t}_2$} \Text(93,91)[]{$\tilde{t}_1$}
\Text(93,29)[]{$H_{i}$} \Text(55,10)[]{$(a)$}
\end{picture}
\hspace{0.8cm}
\begin{picture}(110,110)(0,0)
\DashLine(15,60)(45,60){3} \DashLine(45,60)(70,77.5){3}
\DashLine(70,77.5)(70,42.5){3} \DashLine(70,42.5)(45,60){3}
\DashLine(70,77.5)(95,95){3} \DashLine(70,42.5)(95,25){3}
\Vertex(45,60){1} \Vertex(70,77.5){1} \Vertex(70,42.5){1}
\Text(9,66)[]{$\tilde{t}_2$} \Text(55,80)[]{$H_k$}
\Text(55,42)[]{$\tilde{q}_j$} \Text(78,60)[]{$\tilde{q}_l$}
\Text(103,96)[]{$\tilde{t}_1$} \Text(103,24)[]{$H_{i}$}
\Text(55,10)[]{$(b)$}
\end{picture}
\hspace{0.8cm}
\begin{picture}(110,110)(0,0)
\DashLine(15,60)(45,60){3} \Line(45,60)(70,77.5)
\ArrowLine(70,42.5)(70,77.5) \ArrowLine(45,60)(70,42.5)
\DashLine(70,77.5)(95,95){3} \DashLine(70,42.5)(95,25){3}
\Vertex(45,60){1} \Vertex(70,77.5){1} \Vertex(70,42.5){1}
\Text(9,66)[]{$\tilde{t}_2$}
\Text(55,80)[]{$\tilde{\chi}_k^0(\tilde{\chi}^+_k)$}
\Text(55,45)[]{$t(b)$} \Text(78,60)[]{$t(b)$}
\Text(103,96)[]{$\tilde{t}_1$} \Text(103,24)[]{$H_i$}
\Text(55,10)[]{$(c)$}
\end{picture}
\end{center}

\begin{center}
\begin{picture}(110,110)(0,0)
\DashLine(20,60)(55,60){3} \DashCArc(63.5,68.5)(12,0,360){3}
\DashLine(72,77)(85,90){3} \DashLine(55,60)(85,30){3}
\Vertex(55,60){1} \Vertex(72,77){1} \Text(15,66)[]{$\tilde{t}_2$}
\Text(93,91)[]{$\tilde{t}_1$} \Text(93,29)[]{$H_i$}
\Text(43,80)[]{$H_k$} \Text(83,63)[]{$\tilde{q}_j$}
\Text(55,10)[]{$(d)$}
\end{picture}
\hspace{1.0cm}
\begin{picture}(110,110)(0,0)
\DashLine(12,60)(31,60){3} \DashCArc(43,60)(12,0,360){3}
\DashLine(55,60)(85,90){3} \DashLine(55,60)(85,30){3}
\Vertex(55,60){1} \Vertex(31,60){1} \Text(7,66)[]{$\tilde{t}_2$}
\Text(93,91)[]{$\tilde{t}_1$} \Text(93,29)[]{$H_i$}
\Text(43,80)[]{$H_k$} \Text(43,40)[]{$\tilde{q}_j$}
\Text(55,10)[]{$(e)$}
\end{picture}
\hspace{0.8cm}
\begin{picture}(110,110)(0,0)
\DashLine(20,60)(55,60){3} \DashLine(55,60)(85,90){3}
\DashCArc(63.5,51.5)(12,0,360){3} \DashLine(72,43)(85,30){3}
\Vertex(55,60){1} \Vertex(72,43){1} \Text(15,66)[]{$\tilde{t}_2$}
\Text(93,91)[]{$\tilde{t}_1$} \Text(93,29)[]{$H_i$}
\Text(45,42)[]{$\tilde{q}_l$} \Text(83,63)[]{$\tilde{q}_j$}
\Text(55,10)[]{$(f)$}
\end{picture}
\end{center}
\caption{Feynman diagrams contributing to supersymmetric
electroweak corrections to $\tilde{t}_2\rightarrow \tilde{t}_1
H_i$:\hspace{0.3cm}$H_i$,\hspace{0.1cm}i=1,2,3\hspace{0.1cm}correspond
to $h^0,H^0,A^0$. $(a)$ is tree level diagram; $(b)-(f)$ are
one-loop vertex corrections. In diagram $(b)$ $q=t$ for $k=1...4$
and $q=b$ for $k=5,6$. In diagram $(d)$ and $(e)$ $q=t$ for
$k=1,2$ and $q=b$ for $k=5,6$. In diagram $(f)$,
$q=b,t$}\label{vertex}
\end{figure}
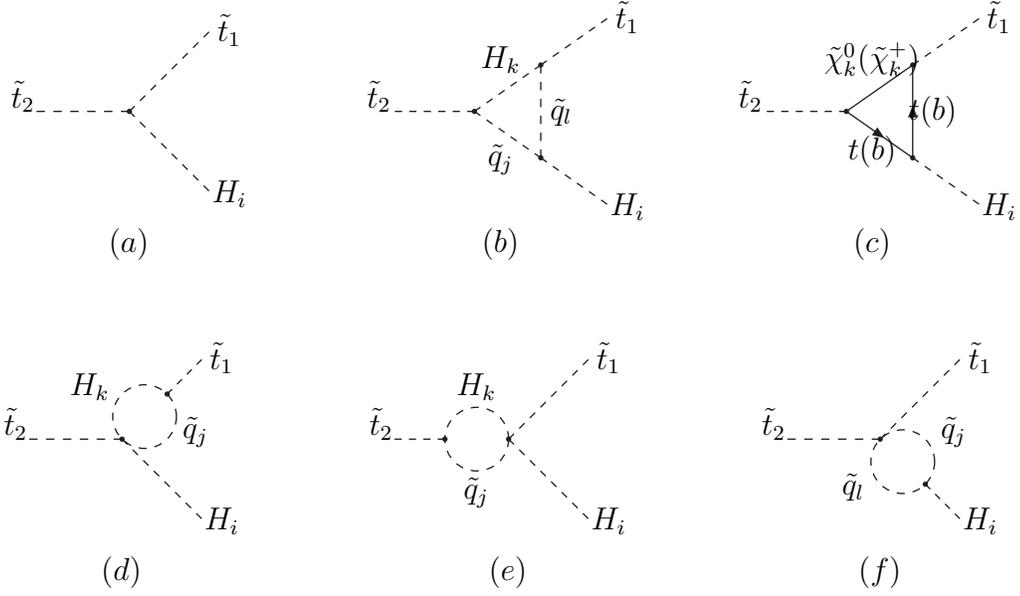
\newpage
\begin{figure}
\begin{center}
\begin{picture}(100,100)(0,0)
\Photon(10,50)(34,50){2}{3} \ArrowArc(50,50)(16,0,180)
\ArrowArc(50,50)(16,180,360) \Photon(66,50)(90,50){2}{3}
\Vertex(34,50){1} \Vertex(66,50){1} \Text(7,40)[]{$Z^0$}
\Text(50,76)[]{$t(b)$} \Text(50,24)[]{$t(b)$}
\Text(93,40)[]{$Z^0$} \Text(50,5)[]{$(a)$}
\end{picture}
\hspace{1.0cm}
\begin{picture}(100,100)(0,0)
\Photon(10,50)(34,50){2}{3} \ArrowArc(50,50)(16,0,180)
\ArrowArc(50,50)(16,180,360) \Photon(66,50)(90,50){2}{3}
\Vertex(34,50){1} \Vertex(66,50){1} \Text(10,40)[]{$W^-$}
\Text(50,76)[]{$t$} \Text(50,24)[]{$b$} \Text(93,40)[]{$W^-$}
\Text(50,5)[]{$(b)$}
\end{picture}
\hspace{1.0cm}
\begin{picture}(100,100)(0,0)
\DashLine(10,50)(34,50){3} \ArrowArc(50,50)(16,0,180)
\ArrowArc(50,50)(16,180,360) \DashLine(66,50)(90,50){3}
\Vertex(34,50){1} \Vertex(66,50){1} \Text(10,40)[]{$H^-$}
\Text(50,76)[]{$t$} \Text(50,24)[]{$b$} \Text(93,40)[]{$H^-(G^-)$}
\Text(50,5)[]{$(c)$}
\end{picture}
\end{center}

\begin{center}
\begin{picture}(100,100)(0,0)
\DashLine(10,50)(34,50){3} \DashCArc(50,50)(16,0,360){3}
\DashLine(66,50)(90,50){3} \Vertex(34,50){1} \Vertex(66,50){1}
\Text(10,40)[]{$H^-$} \Text(50,76)[]{$\tilde{t}_i$}
\Text(50,24)[]{$\tilde{b}_j$} \Text(93,40)[]{$H^-(G^-)$}
\Text(50,5)[]{$(d)$}
\end{picture}
\hspace{1.0cm}
\begin{picture}(100,100)(0,0)
\DashLine(10,40)(90,40){3} \DashCArc(50,54)(14,0,360){3}
\Vertex(50,40){1} \Text(10,30)[]{$H^-$}
\Text(50,80)[]{$\tilde{t}_i,\tilde{b}_i$}
\Text(93,30)[]{$H^-(G^-)$} \Text(50,5)[]{$(e)$}
\end{picture}
\hspace{1.0cm}
\begin{picture}(100,100)(0,0)
\DashLine(10,50)(34,50){3} \ArrowArc(50,50)(16,0,180)
\ArrowArc(50,50)(16,180,360) \DashLine(66,50)(90,50){3}
\Vertex(34,50){1} \Vertex(66,50){1} \Text(10,40)[]{$H_i$}
\Text(50,76)[]{$t$} \Text(50,24)[]{$b$} \Text(93,40)[]{$H_j$}
\Text(50,5)[]{$(f)$}
\end{picture}
\end{center}

\begin{center}
\begin{picture}(100,100)(0,0)
\DashLine(10,50)(34,50){3} \DashCArc(50,50)(16,0,360){3}
\DashLine(66,50)(90,50){3} \Vertex(34,50){1} \Vertex(66,50){1}
\Text(10,40)[]{$H_i$} \Text(50,76)[]{$\tilde{t}_k$}
\Text(50,24)[]{$\tilde{b}_l$} \Text(93,40)[]{$H_j$}
\Text(50,5)[]{$(g)$}
\end{picture}
\hspace{1.0cm}
\begin{picture}(100,100)(0,0)
\DashLine(10,40)(90,40){3} \DashCArc(50,54)(14,0,360){3}
\Vertex(50,40){1} \Text(10,30)[]{$H_i$}
\Text(50,80)[]{$\tilde{t}_k,\tilde{b}_l$} \Text(93,30)[]{$H_j$}
\Text(50,5)[]{$(h)$}
\end{picture}
\hspace{1.0cm}
\begin{picture}(100,100)(0,0)
\DashLine(10,50)(34,50){3} \DashCArc(50,50)(16,0,360){3}
\DashLine(66,50)(90,50){3} \Vertex(34,50){1} \Vertex(66,50){1}
\Text(10,40)[]{$\tilde{t}_i(\tilde{b}_i)$} \Text(50,76)[]{$H_k$}
\Text(50,24)[]{$\tilde{q}_l$}
\Text(93,40)[]{$\tilde{t}_j(\tilde{b}_j)$} \Text(50,5)[]{$(i)$}
\end{picture}
\end{center}

\begin{center}
\begin{picture}(100,100)(0,0)
\DashLine(10,50)(34,50){3} \CArc(50,50)(16,0,360)
\DashLine(66,50)(90,50){3} \Vertex(34,50){1} \Vertex(66,50){1}
\Text(10,40)[]{$\tilde{t}_i(\tilde{b}_i)$}
\Text(50,76)[]{$\tilde{\chi}^0_k$} \Text(50,24)[]{$t(b)$}
\Text(93,40)[]{$\tilde{t}_j(\tilde{b}_j)$} \Text(50,5)[]{$(j)$}
\end{picture}
\hspace{1.0cm}
\begin{picture}(100,100)(0,0)
\DashLine(10,50)(34,50){3} \CArc(50,50)(16,0,360)
\DashLine(66,50)(90,50){3} \Vertex(34,50){1} \Vertex(66,50){1}
\Text(10,40)[]{$\tilde{t}_i(\tilde{b}_i)$}
\Text(50,76)[]{$\tilde{\chi}^+_k$} \Text(50,24)[]{$b(t)$}
\Text(93,40)[]{$\tilde{t}_j(\tilde{b}_j)$} \Text(50,5)[]{$(k)$}
\end{picture}
\hspace{1.0cm}
\begin{picture}(100,100)(0,0)
\DashLine(10,40)(90,40){3} \DashCArc(50,54)(14,0,360){3}
\Vertex(50,40){1} \Text(10,30)[]{$\tilde{t}_i(\tilde{b}_i)$}
\Text(50,80)[]{$H_k,\tilde{t}_l,\tilde{b}_l$}
\Text(93,30)[]{$\tilde{t}_j(\tilde{b}_j)$} \Text(50,5)[]{$(l)$}
\end{picture}
\end{center}

\begin{center}
\begin{picture}(100,100)(0,0)
\ArrowLine(10,50)(34,50) \ArrowLine(34,50)(66,50)
\ArrowLine(66,50)(90,50) \DashCArc(50,50)(16,0,180){3}
\Vertex(34,50){1} \Vertex(66,50){1} \Text(10,40)[]{$t(b)$}
\Text(50,40)[]{$q$} \Text(50,75)[]{$H_k$} \Text(93,40)[]{$t(b)$}
\Text(50,5)[]{$(m)$}
\end{picture}
\hspace{1.0cm}
\begin{picture}(100,100)(0,0)
\ArrowLine(10,50)(34,50) \Line(34,50)(66,50)
\ArrowLine(66,50)(90,50) \DashCArc(50,50)(16,0,180){3}
\Vertex(34,50){1} \Vertex(66,50){1} \Text(10,40)[]{$t(b)$}
\Text(50,40)[]{$\tilde{\chi}^0_k$}
\Text(50,75)[]{$\tilde{t}_i(\tilde{b}_i)$} \Text(93,40)[]{$t(b)$}
\Text(50,5)[]{$(n)$}
\end{picture}
\hspace{1.0cm}
\begin{picture}(100,100)(0,0)
\ArrowLine(10,50)(34,50) \Line(34,50)(66,50)
\ArrowLine(66,50)(90,50) \DashCArc(50,50)(16,0,180){3}
\Vertex(34,50){1} \Vertex(66,50){1} \Text(10,40)[]{$t(b)$}
\Text(50,40)[]{$\tilde{\chi}^+_k$}
\Text(50,75)[]{$\tilde{b}_i(\tilde{t}_i)$} \Text(93,40)[]{$t(b)$}
\Text(50,5)[]{$(o)$}
\end{picture}
\end{center}

\begin{center}
\begin{picture}(100,100)(0,0)
\ArrowLine(10,50)(34,50) \Line(34,50)(66,50)
\ArrowLine(66,50)(90,50) \DashCArc(50,50)(16,0,180){3}
\Vertex(34,50){1} \Vertex(66,50){1}
\Text(10,40)[]{$\tilde{\chi}^+_i$} \Text(50,40)[]{$t(b)$}
\Text(50,75)[]{$\tilde{b}_k(\tilde{t}_k)$}
\Text(93,40)[]{$\tilde{\chi}^+_j$} \Text(50,5)[]{$(p)$}
\end{picture}
\hspace{1.0cm}
\begin{picture}(100,100)(0,0)
\DashLine(50,55)(50,30){3} \ArrowArc(50,70)(15,0,180)
\CArc(50,70)(15,180,360) \Vertex(50,55){1}
\Text(75,40)[]{$H^0,h^0$} \Text(50,95)[]{$t,b$}
\Text(50,5)[]{$(q)$}
\end{picture}
\hspace{1.0cm}
\begin{picture}(100,100)(0,0)
\DashLine(50,55)(50,30){3} \DashCArc(50,70)(15,0,360){3}
\Vertex(50,55){1} \Text(75,40)[]{$H^0,h^0$}
\Text(50,95)[]{$\tilde{t}_i,\tilde{b}_i$} \Text(50,5)[]{$(r)$}
\end{picture}
\end{center}
\caption{Feynman diagrams contributing to renormalization
constants. In diagram $(i)$ $q=t(b)$ for  $k=1...4$ and $q=b(t)$
for $k=5,6$. In diagram $(f)$ $(g)$ $(h)$ $i=j=1,2,3$ or
$i=1,j=2$} \label{self}
\end{figure}
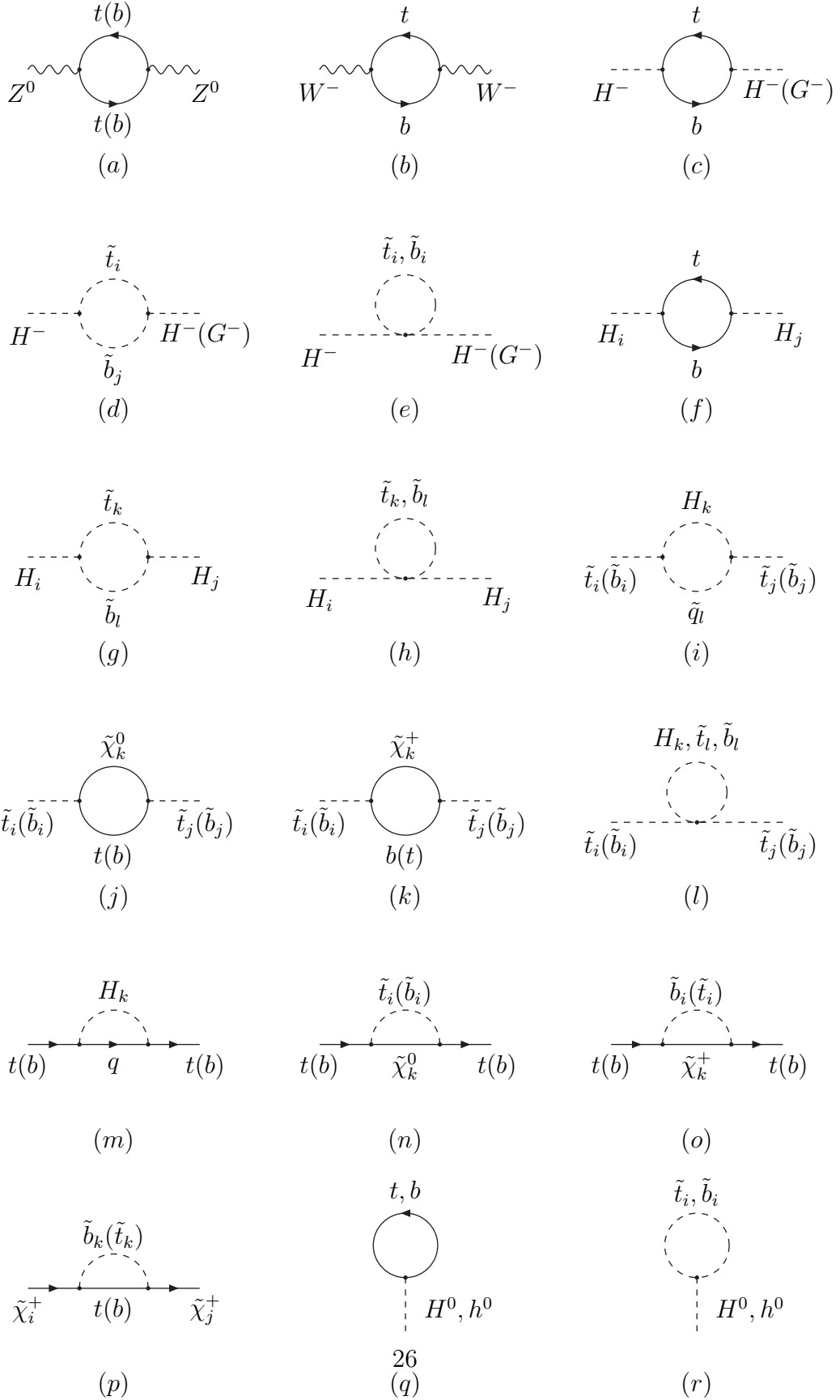
\newpage
\begin{figure}
\epsfxsize=15 cm \centerline{ \epsffile{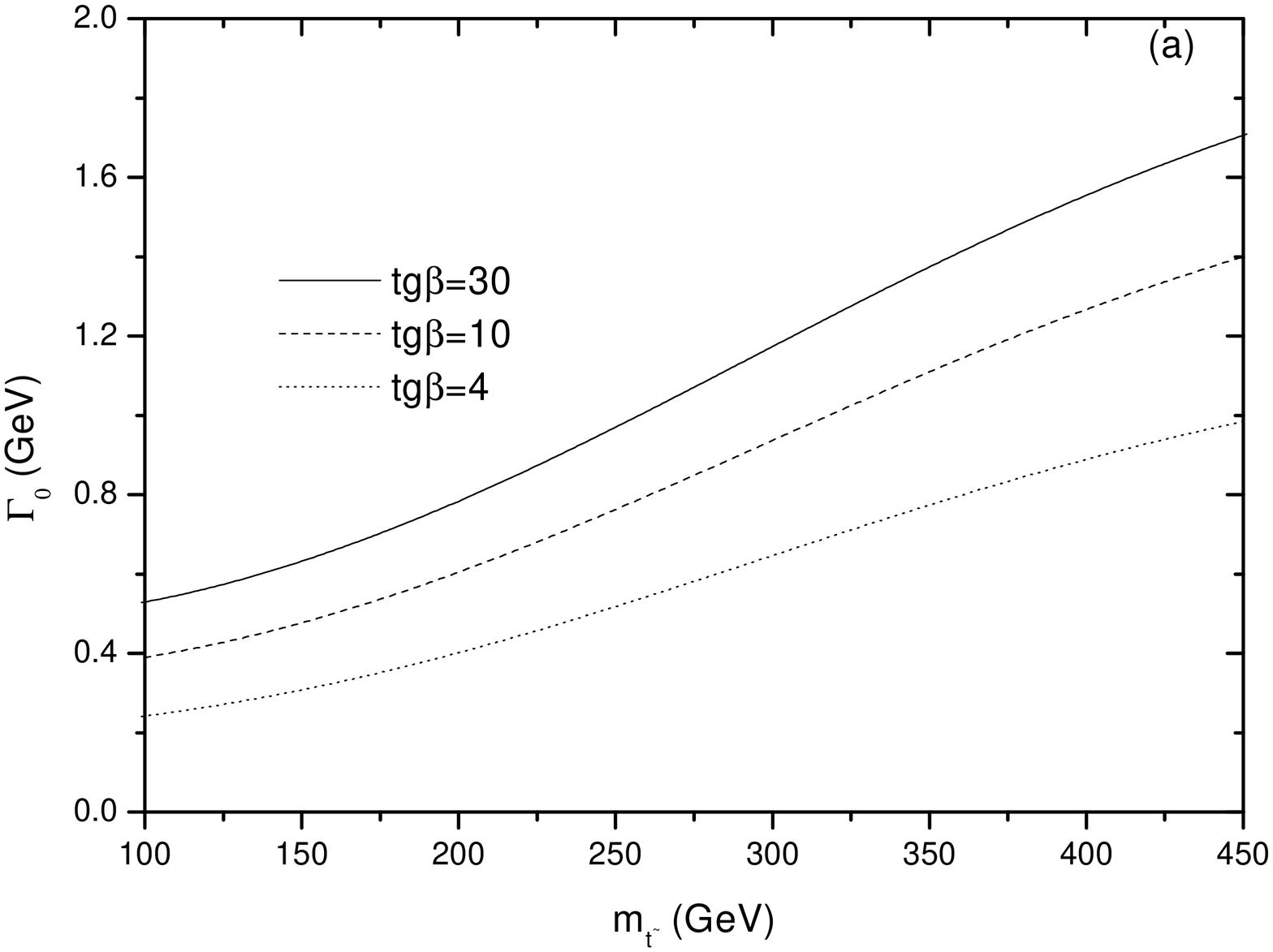 }}
 \epsfxsize=15
cm \centerline{ \epsffile{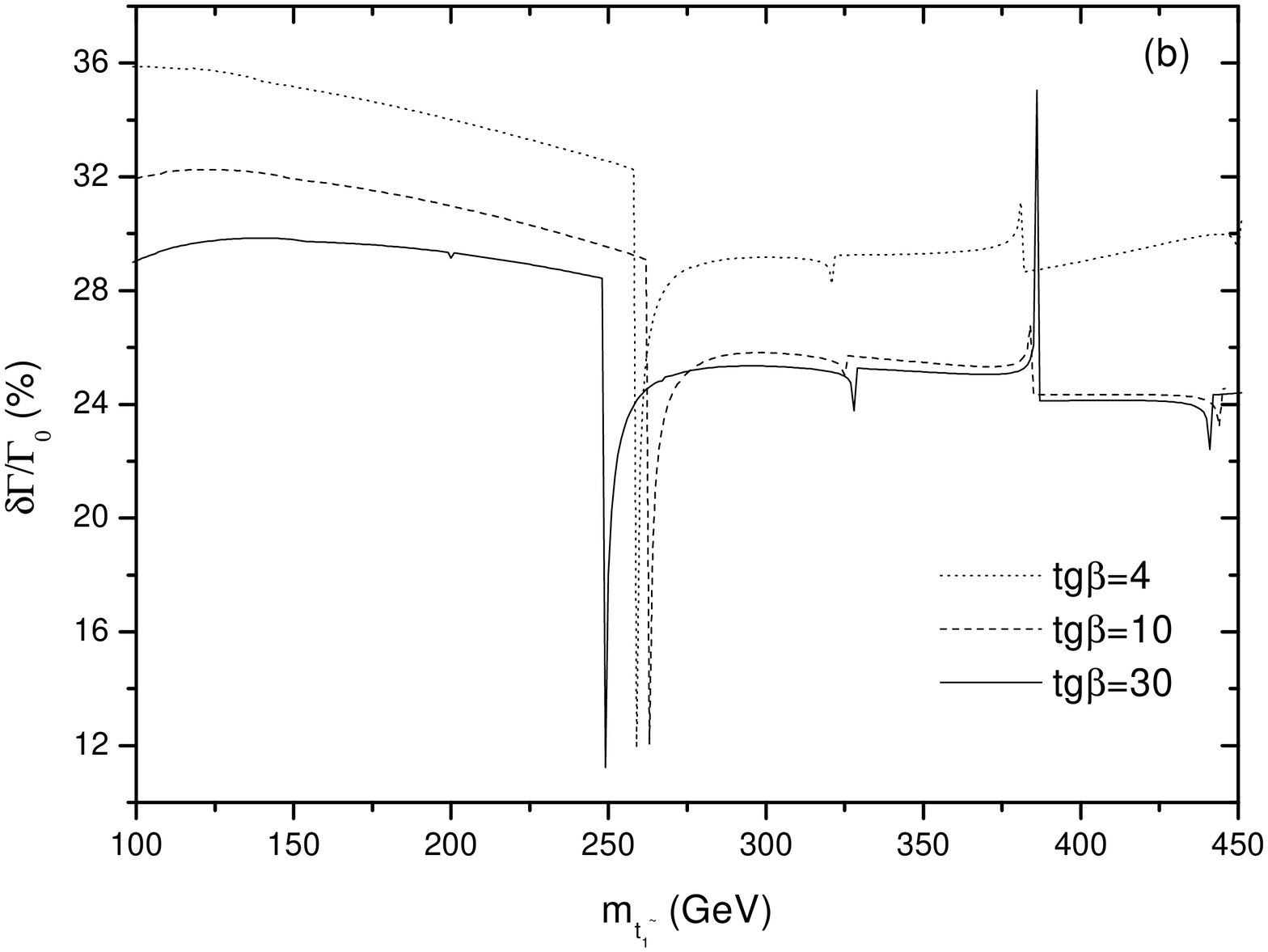 }}
 \caption{The
tree-level decay width  $(a)$ of $\tilde{t}_2 \rightarrow
\tilde{t}_1h^0$ and its Yukawa corrections $(b)$ as functions of
$m_{\tilde{t}_1}$ for $\tan\beta=4$, $10,$  and $30$,
respectively, assuming $m_{A^0}=150$GeV, $\mu=M_2=200$GeV,
$A_t=A_b=600$GeV and
$M_{\tilde Q}=1.5M_{\tilde U}=1.5M_{\tilde D}$.} 
\end{figure}
\begin{figure}
\centerline{\epsfig{file=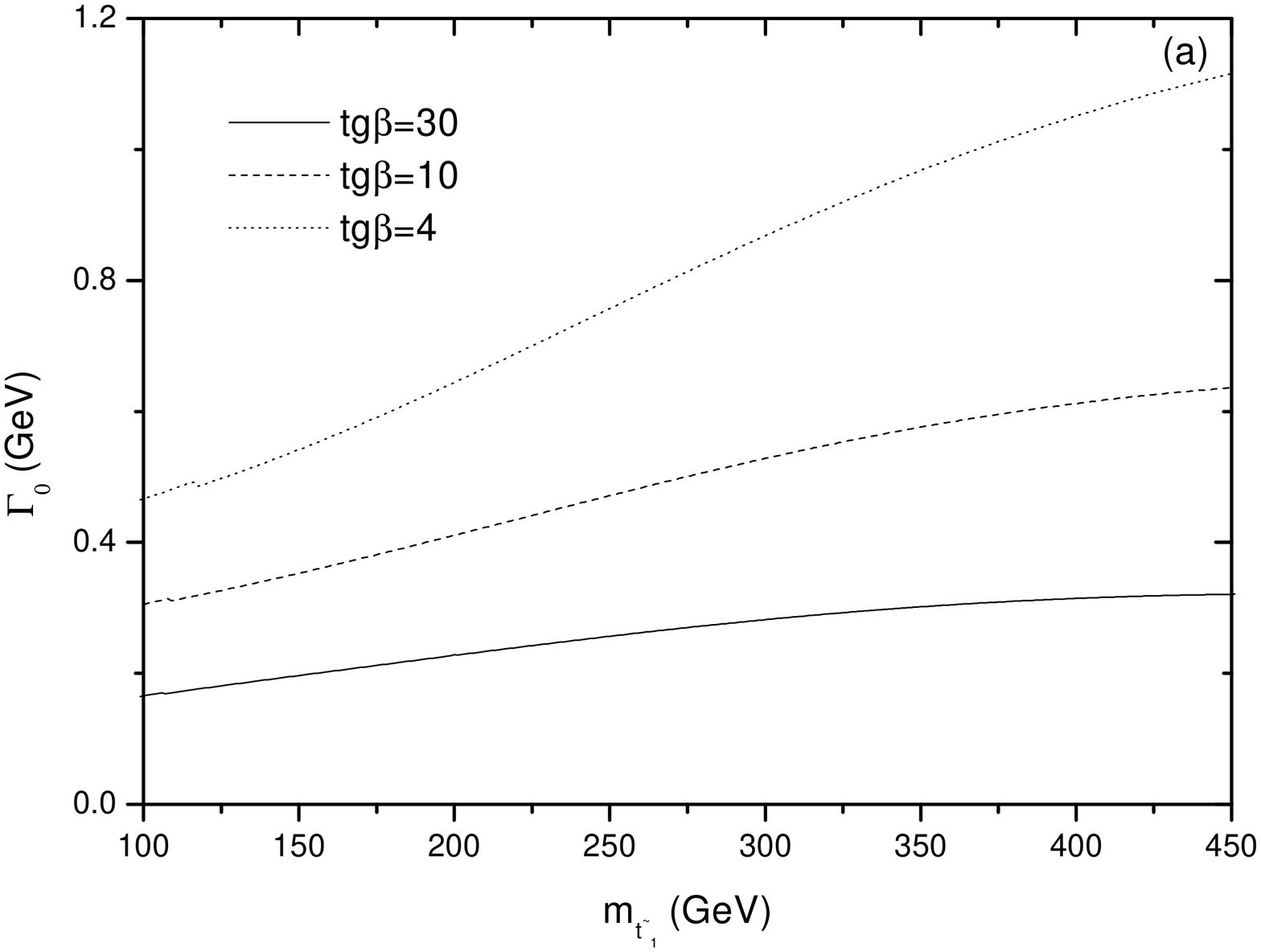, width=400pt}}
\centerline{\epsfig{file=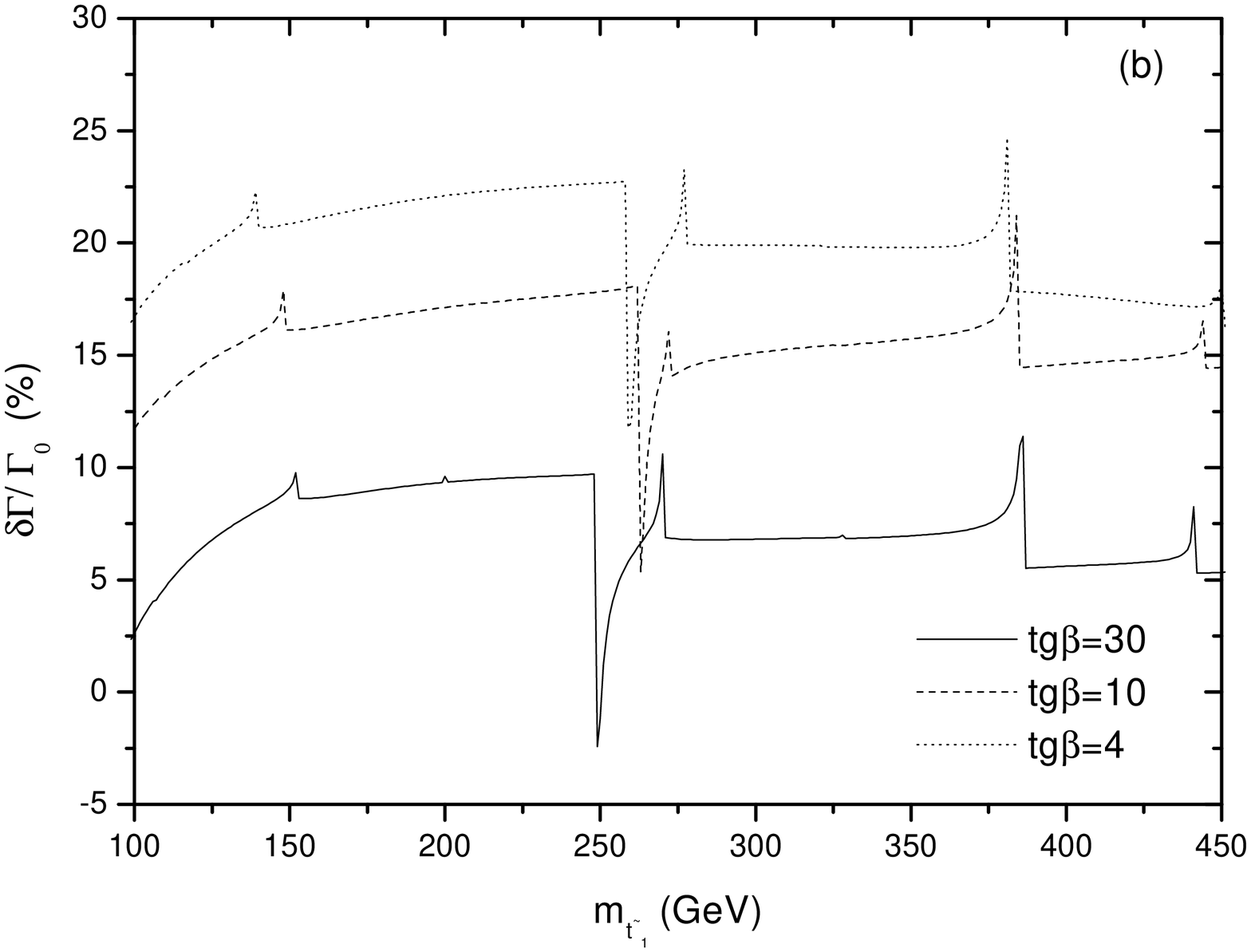, width=400pt}} \caption{The
tree-level decay width $(a)$ of $\tilde{t}_2 \rightarrow
\tilde{t}_1H^0$ and its Yukawa corrections $(b)$ as functions of
$m_{\tilde{t}_1}$ for $\tan\beta=4$, 10, and 30, respectively,
assuming $m_{A^0}=150$GeV, $\mu=M_2=200$GeV, $A_t=A_b=600$GeV and
$M_{\tilde Q}=1.5M_{\tilde U}=1.5M_{\tilde D}$.} 
\end{figure}
\begin{figure}
\centerline{\epsfig{file=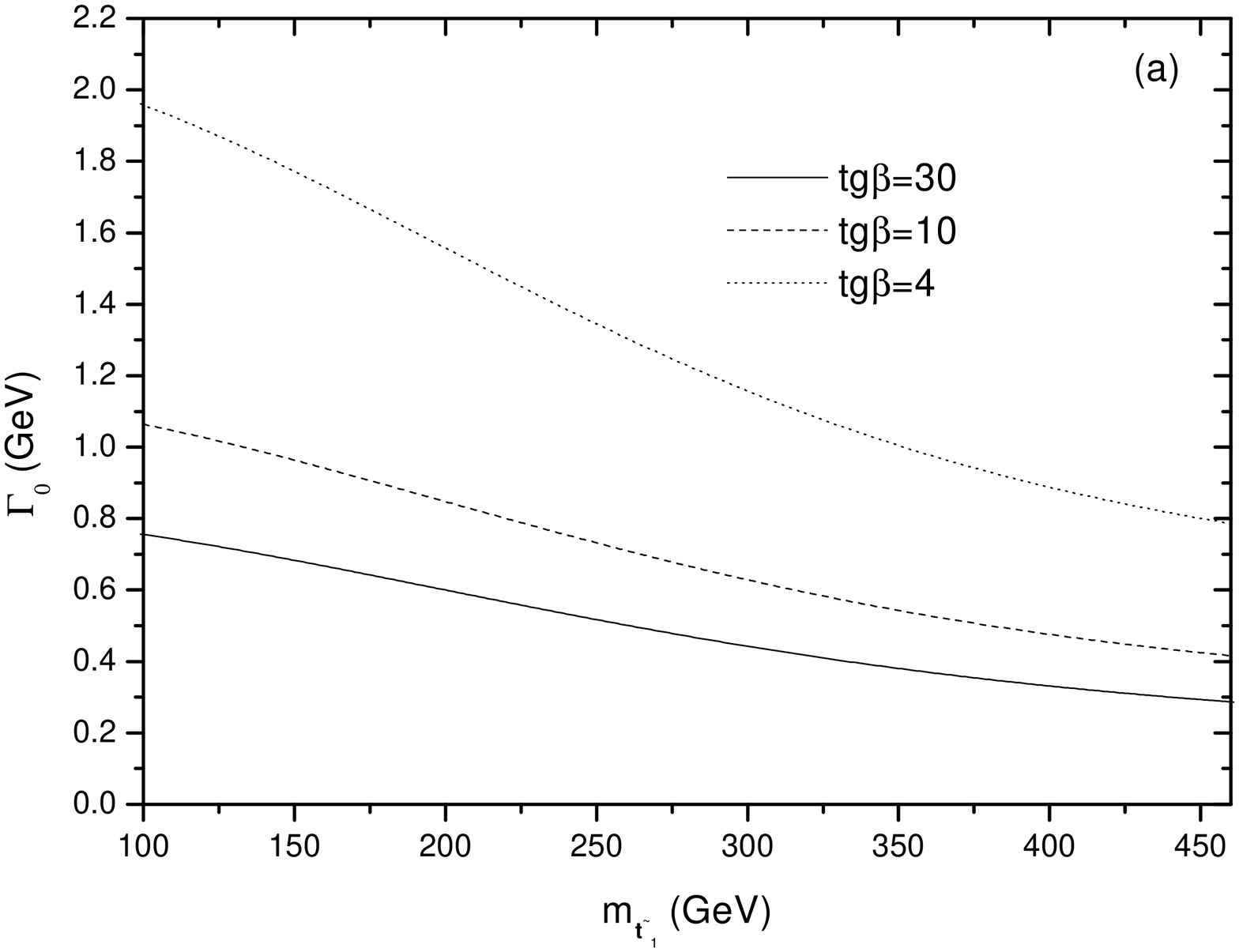,width=400pt}}
\centerline{\epsfig{file=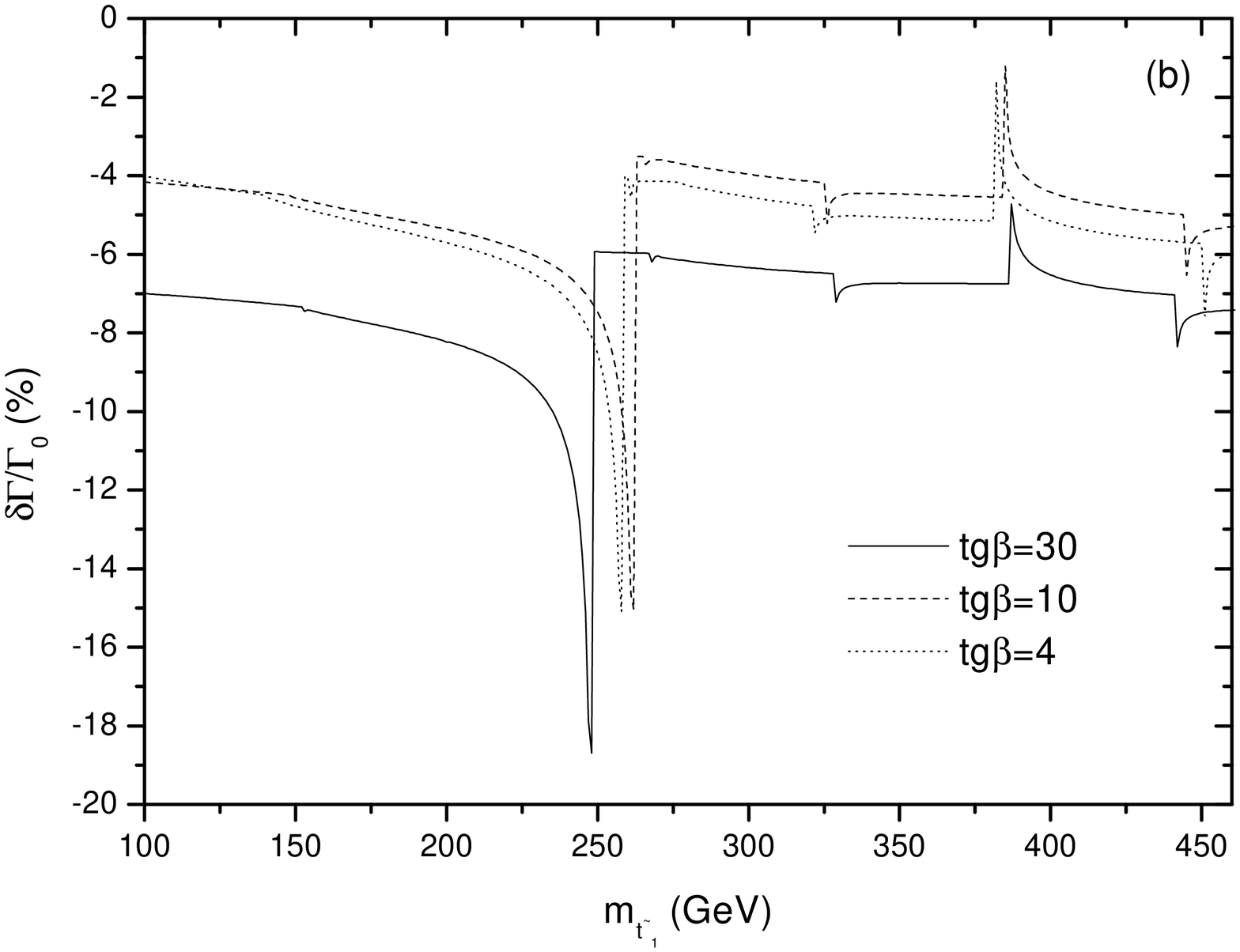, width=400pt}} \caption{The
tree-level decay width  $(a)$ of $\tilde{t}_2 \rightarrow
\tilde{t}_1A^0$ and its Yukawa corrections  $(b)$ as functions of
$m_{\tilde{t}_1}$ for $\tan\beta=4$, $10,$ and $30$, respectively,
assuming $m_{A^0}=150$GeV, $\mu=M_2=200$GeV, $A_t=A_b=600$GeV and
$M_{\tilde Q}=1.5M_{\tilde U}=1.5M_{\tilde D}$.} 
\end{figure}
\begin{figure}
\centerline{\epsfig{file=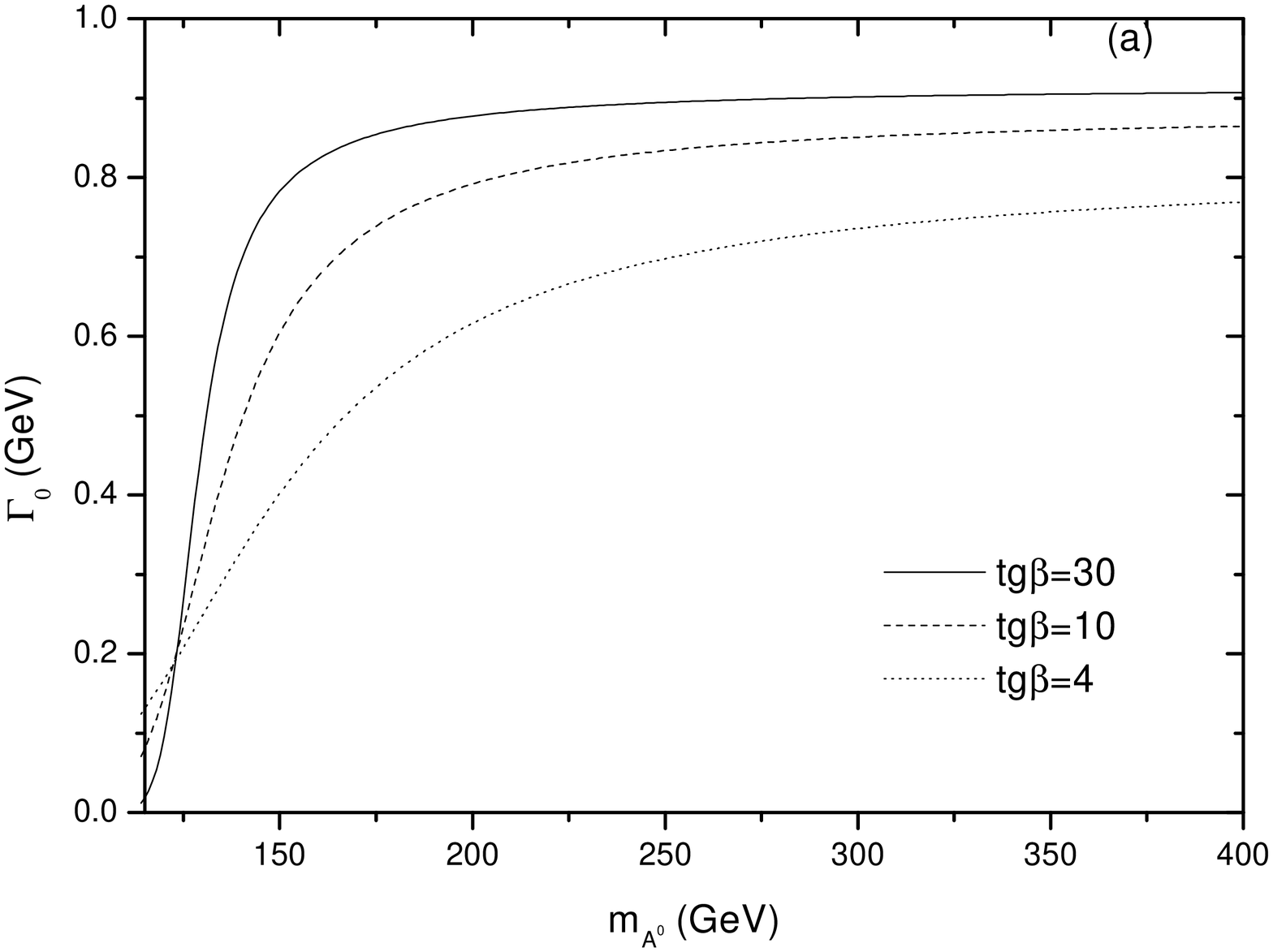,width=400pt}}
\centerline{\epsfig{file=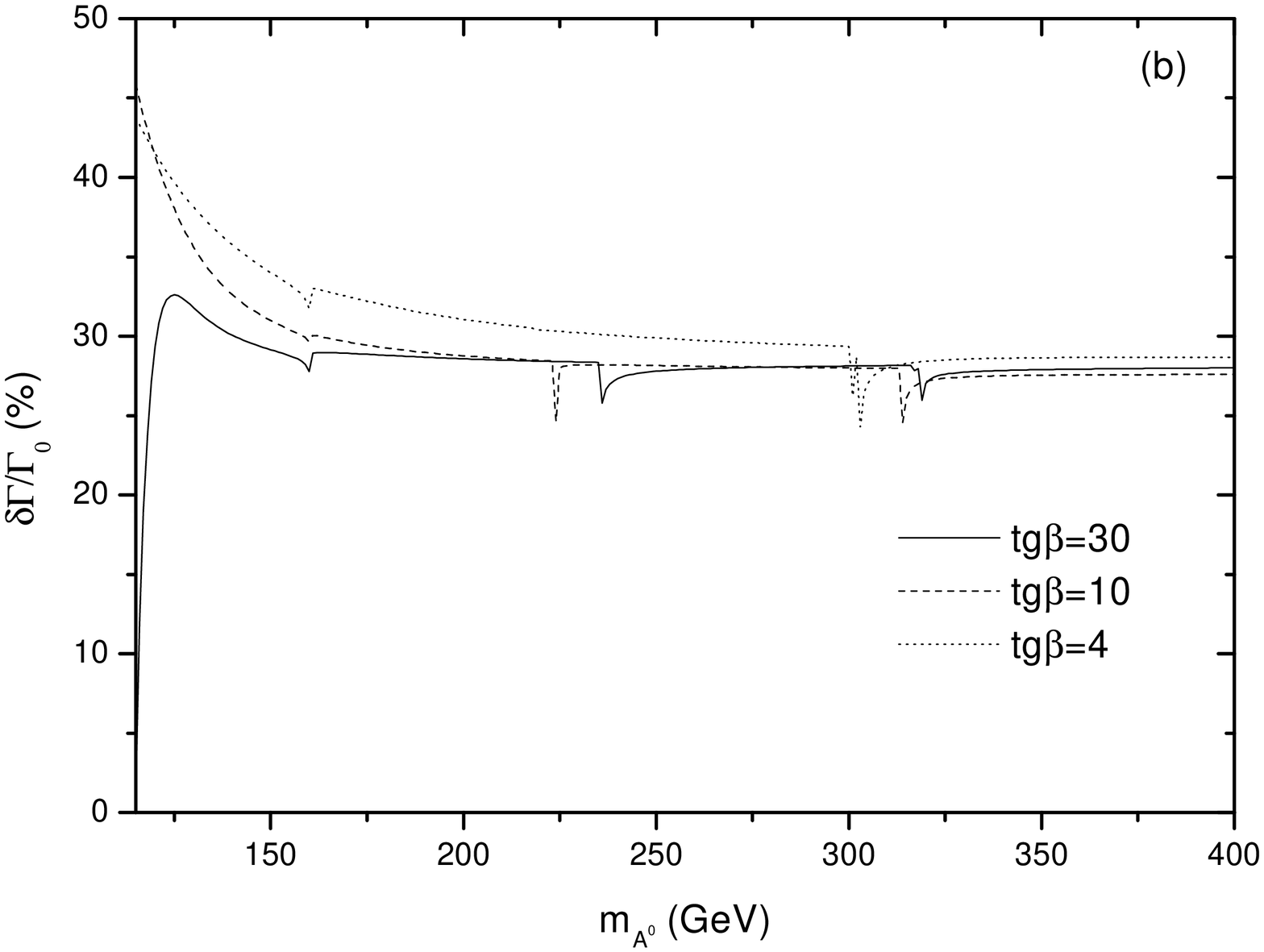, width=400pt}}
 \caption{The tree-level decay width $(a)$ of
$\tilde{t}_2 \rightarrow \tilde{t}_1h^0$ and its Yukawa
corrections  $(b)$ as functions of $m_{A^0}$ for $\tan\beta=4$,
$10,$ and $30$, respectively, assuming $m_{\tilde{t}_1}=200$GeV,
$\mu=M_2=200$GeV, $A_t=A_b=600$GeV and
$M_{\tilde Q}=1.5M_{\tilde U}=1.5M_{\tilde D}$.} 
\end{figure}
\begin{figure}
\psfig{file=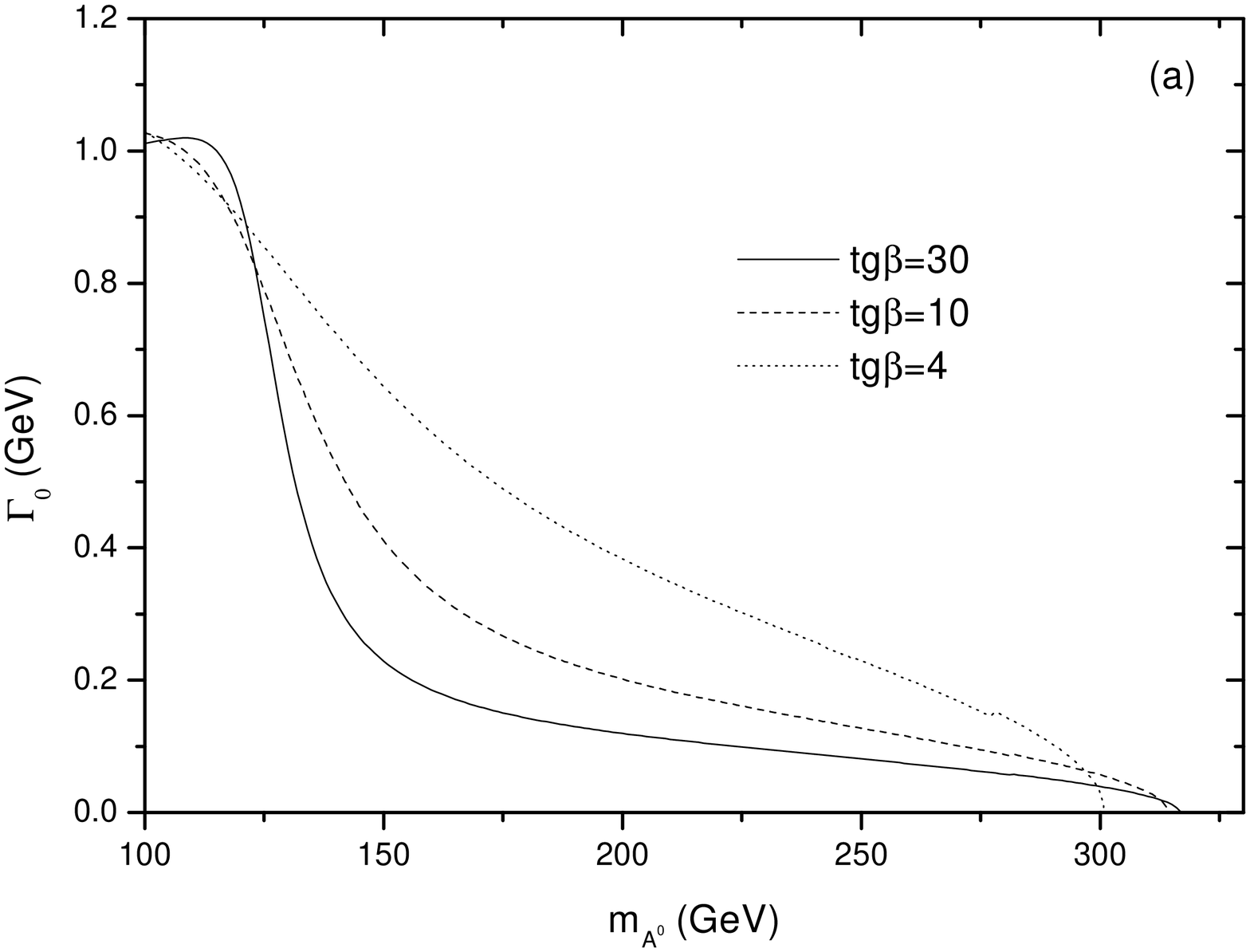, width=400pt} \psfig{file=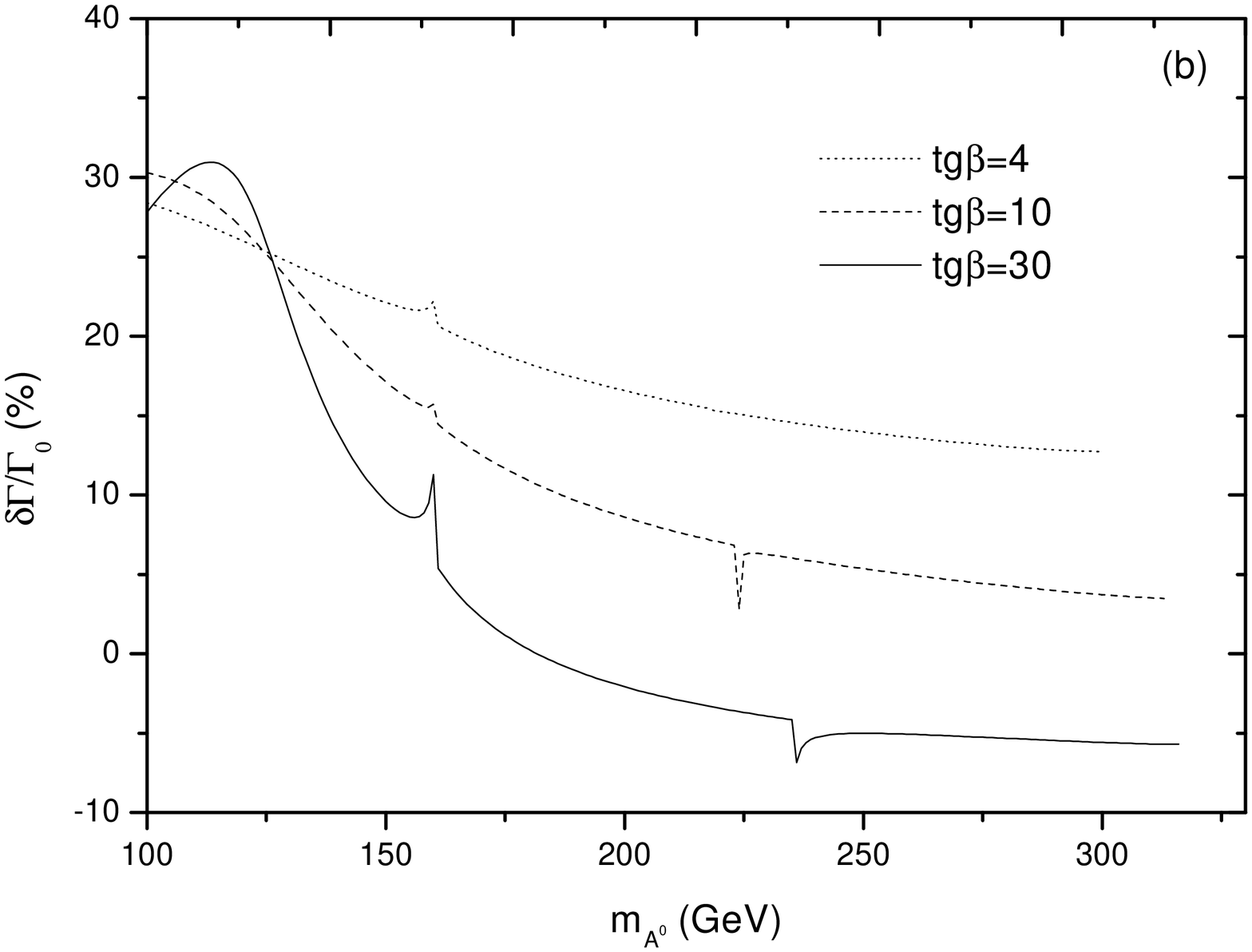,
width=400pt} \caption{The tree-level decay width $(a)$ of
$\tilde{t}_2 \rightarrow \tilde{t}_1H^0$ and its Yukawa
corrections  $(b)$ as functions of $m_{A^0}$ for $\tan\beta=4$,
$10$,  and  $30$, respectively, assuming $m_{\tilde{t}_1}=200$GeV,
$\mu=M_2=200$GeV, $A_t=A_b=600$GeV and
$M_{\tilde Q}=1.5M_{\tilde U}=1.5M_{\tilde D}$.} 
\end{figure}
\begin{figure}
\psfig{file=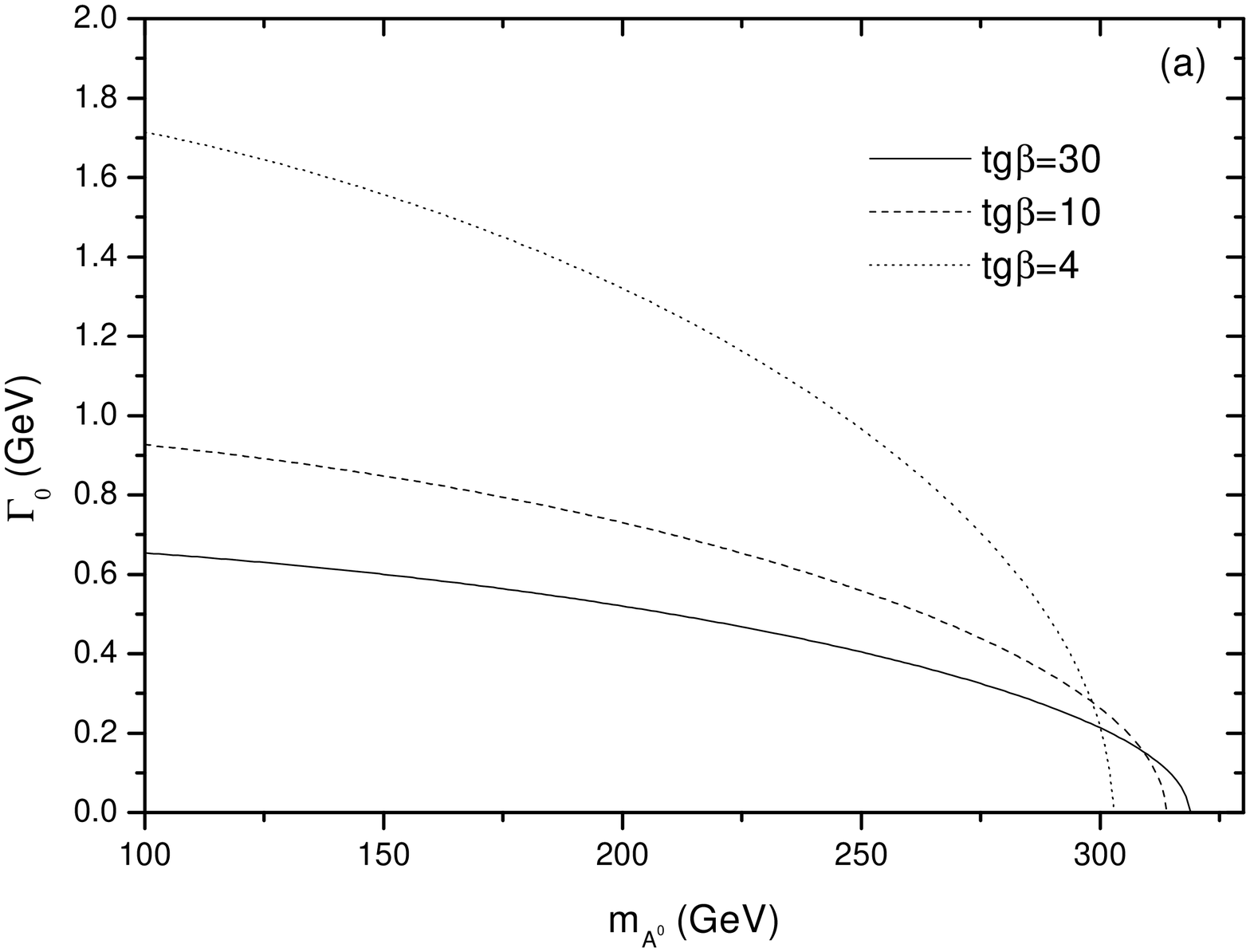, width=400pt} \psfig{file=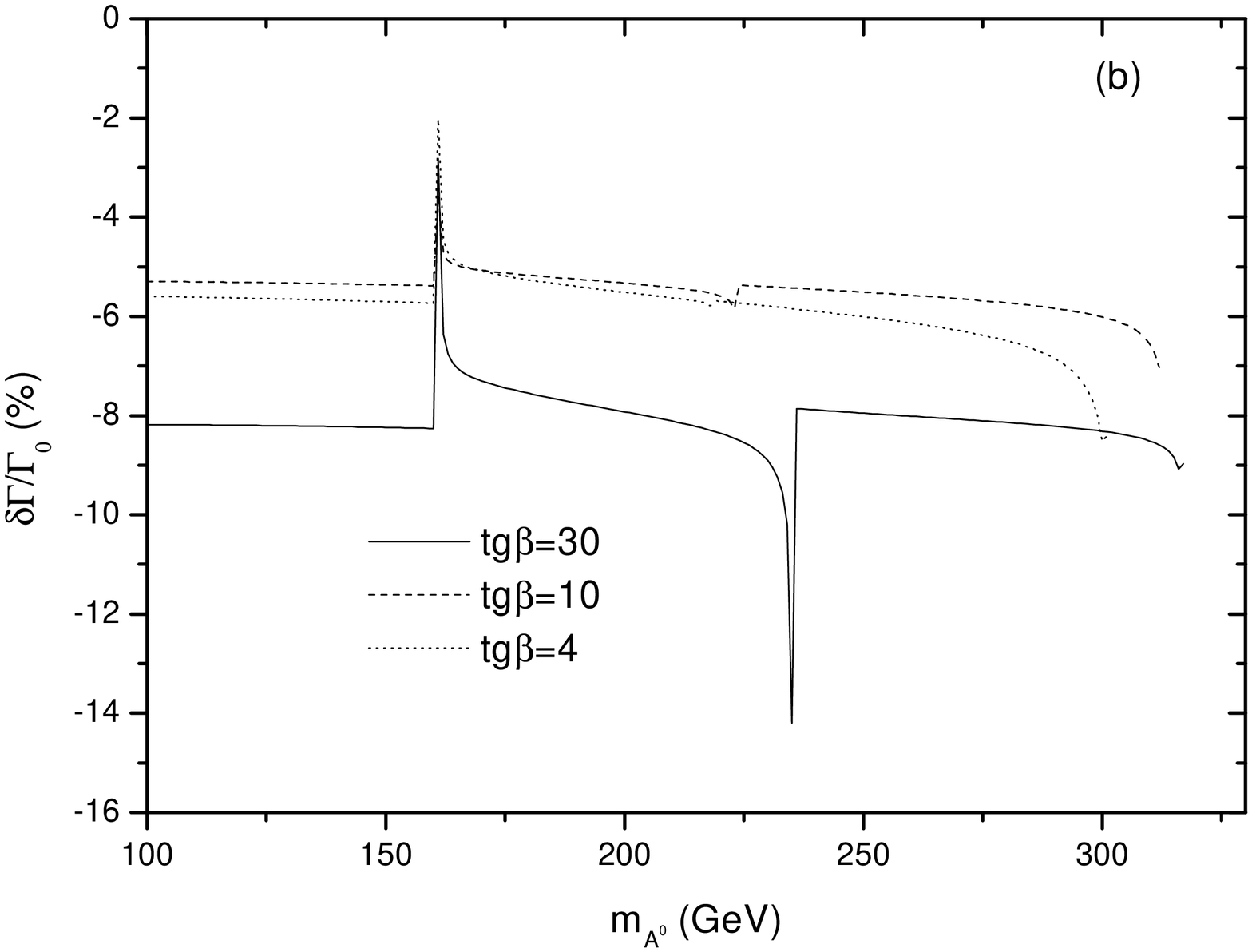,
width=400pt} \caption{The tree-level decay width $(a)$ of
$\tilde{t}_2 \rightarrow \tilde{t}_1A^0$ and its Yukawa
corrections $(b)$ as functions of $m_{A^0}$ for $\tan\beta=4$,
$10,$ and $30$, respectively, assuming $m_{\tilde{t}_1}=200$GeV,
$\mu=M_2=200$GeV, $A_t=A_b=600$GeV, and
$M_{\tilde Q}=1.5M_{\tilde U}=1.5M_{\tilde D}$.} 
\end{figure}
\begin{figure}
\psfig{file=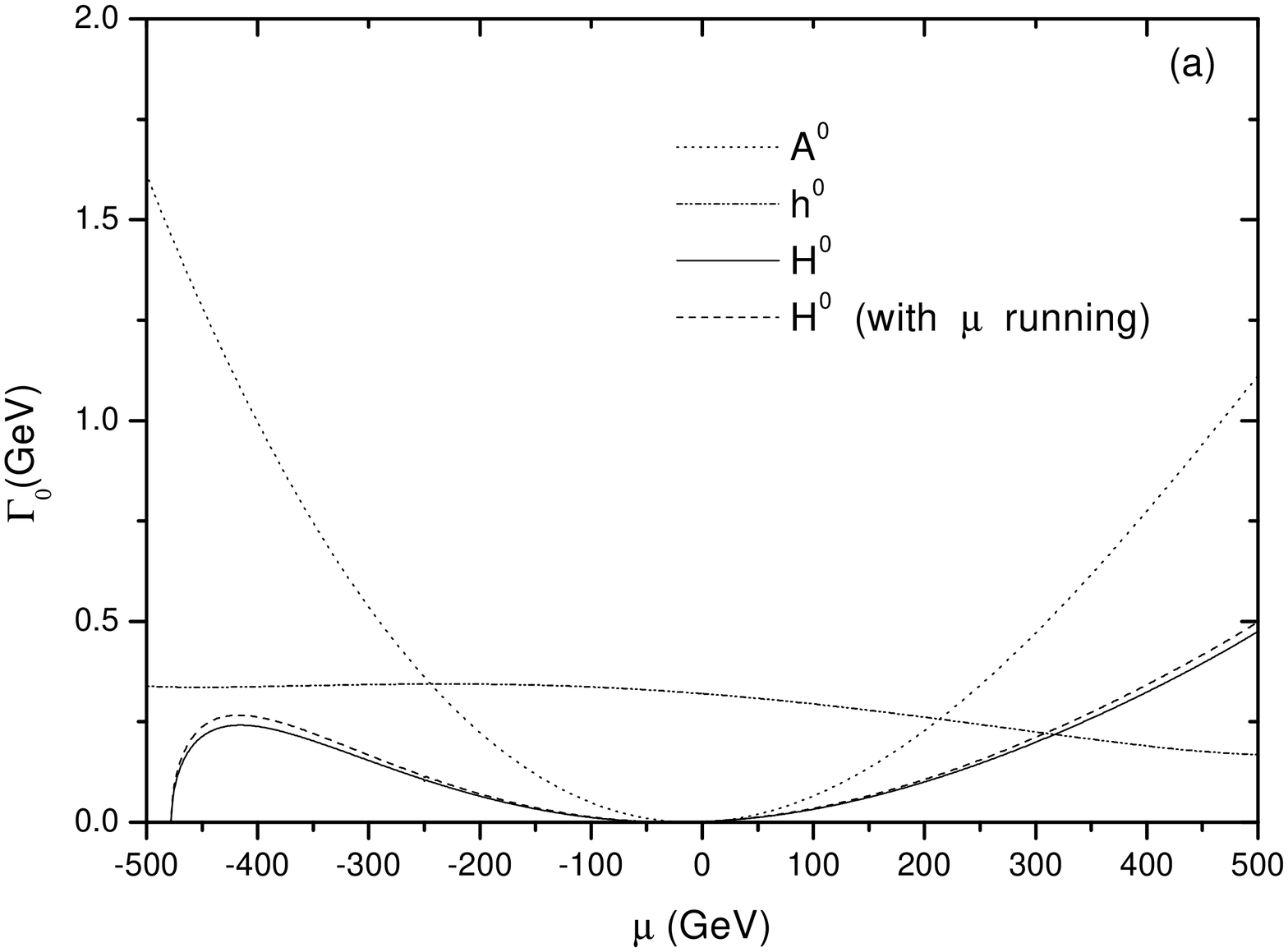,width=400pt}
\psfig{file=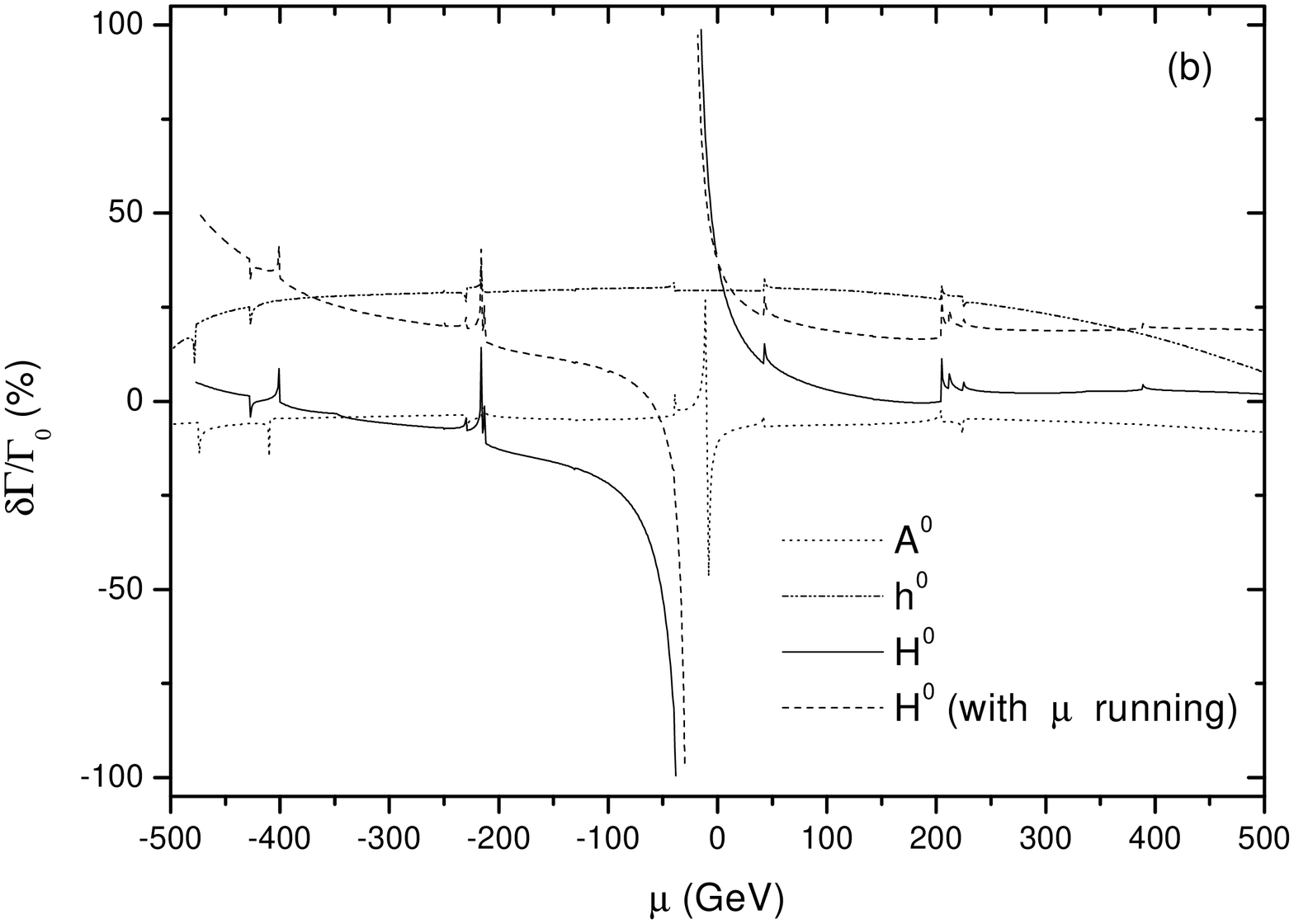,width=400pt} \caption{The tree-level decay
width $(a)$ of $\tilde{t}_2 \rightarrow \tilde{t}_1H^0_i$ and its
Yukawa corrections $(b)$ as functions of $\mu$, assuming
$\tan\beta=30$, $m_{\tilde{t}_1}=250$GeV, $M_2=100$GeV,
$A_t=250GeV$,$A_b=-250$GeV, $m_{A^0}=150$GeV and $M_{\tilde
Q}=1.5M_{\tilde U}=1.5M_{\tilde D}$.} 
\end{figure}
\begin{figure}
\psfig{file=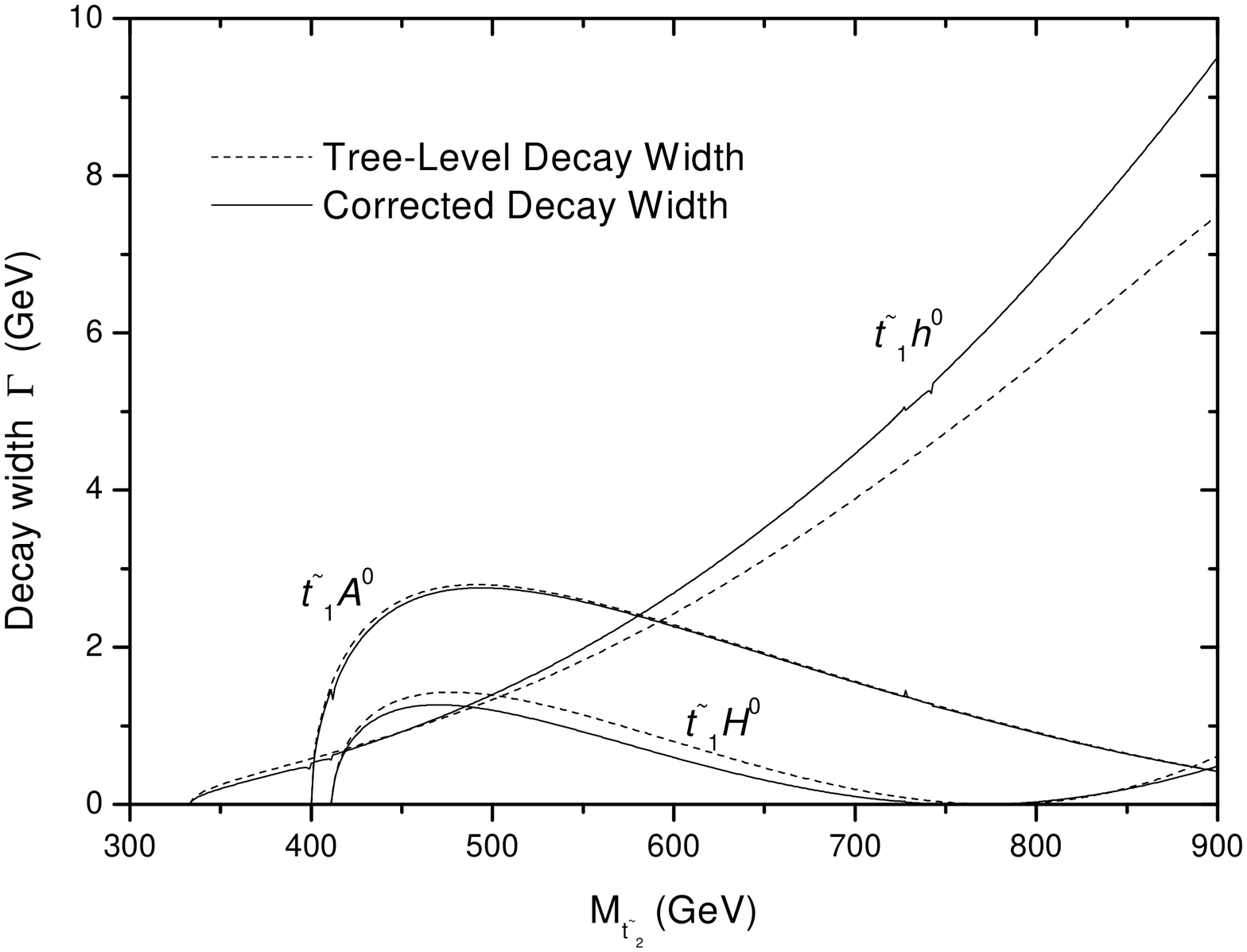,width=400pt} \caption{The decay width of
$\tilde{t}_2 \rightarrow \tilde{t}_1H^0_i$ as a function of
$m_{\tilde{t_2}}$, assuming $\tan\beta=3$,$\cos\theta_{\tilde
t}=0.26$, $m_{\tilde{t}_1}=250$GeV, $\mu=550$GeV,  $m_{\tilde
g}=600$GeV, $A_t=A_b$, $m_{A^0}=150$GeV and $M_{\tilde
D}=1.12M_{\tilde Q}$. The solid lines correspond to the
Yukawa-corrected decay widths, The dashed lines correspond to the
tree-level decay widths. } 
\end{figure}
\begin{figure}
\psfig{file=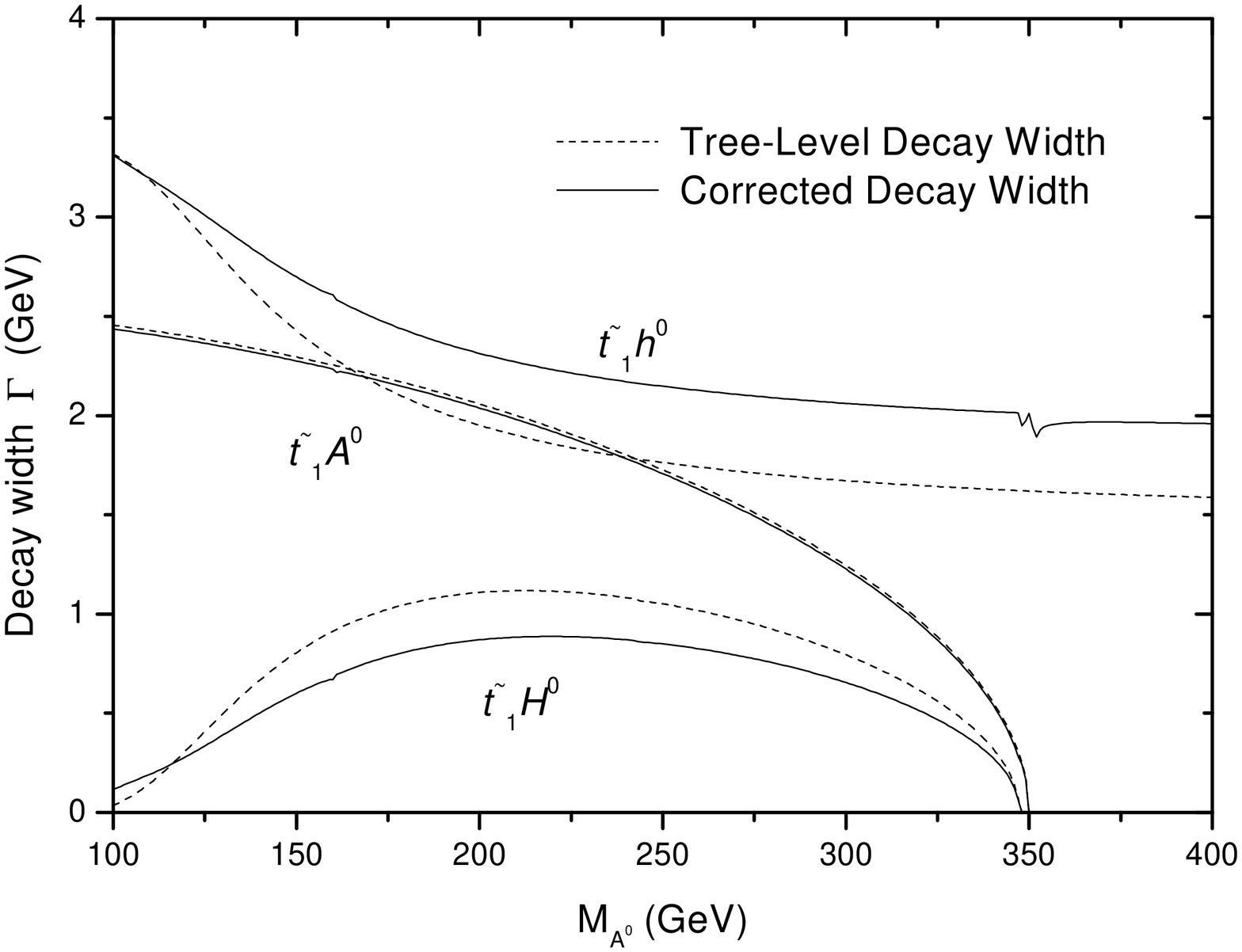,width=400pt} \caption{The decay width of
$\tilde{t}_2 \rightarrow \tilde{t}_1H^0_i$ as a function of
$m_{A^0}$, assuming $\tan\beta=3$,$\cos\theta_{\tilde t}=0.26$,
$m_{\tilde{t}_1}=250$GeV, $\mu=550$GeV, $m_{\tilde g}=600$GeV,
$A_t=A_b$ and $M_{\tilde D}=1.12M_{\tilde Q}$. The solid lines
correspond to the Yukawa-corrected decay widths, The dashed lines
correspond to the tree-level decay widths. } 
\end{figure}
\begin{figure}
\psfig{file=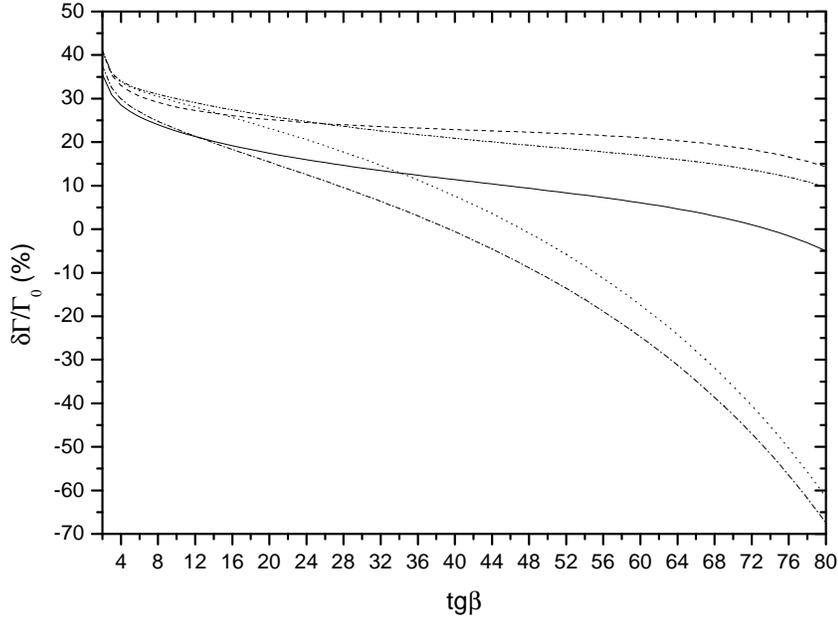, width=400pt} \caption{The Yukawa corrections of
$\tilde{t}_2 \rightarrow \tilde{t}_1H^0$ as a function of
$\tan\beta$, assuming $m_{\tilde{t}_1}=250$GeV, $M_2=200$GeV,
$A_t=A_b=900$GeV, $\mu=200$GeV, $m_{A^0}=150$GeV and $M_{\tilde
Q}=1.5M_{\tilde U}=1.5M_{\tilde D}$. The dotted line corresponds
to the corrections using the on-shell parameters; the dashed line
corresponds to the corrections using the running parameters
$\hat{m}_t(Q)$, $\hat{m}_b(Q)$, $\hat{A}_t$, and $\hat\mu$;  the
solid line corresponds to the corrections using the same running
parameters except the running $\hat\mu$; the dashed-dotted line to
the improved result only using the running mass $\hat{m}_t(Q)$;
and the dash-dot-doted line to the improved
result only using the running mass $\hat{m}_b(Q)$.} 
\end{figure}
\end{document}